\documentclass[
prb,
twocolumn,
superscriptaddress,
a4paper,
showpacs,
showkeys,
floatfix
]{revtex4}
\usepackage[final]{graphicx}
\usepackage{subfigure}
\usepackage{nomencl}
\usepackage{amsmath}

\newcommand{\etal}{\textit{et al.}}
\newcommand{\kB}{k_\mrm{B}}
\newcommand{\dd}{\mathrm{d}}
\newcommand{\kT}{\kB T}
\newcommand{\mrm}[1]{\mathrm{#1}}

\newcommand{\xn}{x_\mrm{N}}

\newcommand{\rc}{r_\mrm{c}}

\newcommand{\istar}{i^*}
\newcommand{\name}{}

\newcommand{\Ln}[1]{\ln\left(#1\right)}
\newcommand{\Exp}[1]{\exp\left[#1\right]}

\newcommand{\Sin}[1]{\sin\left(#1\right)}

\newcommand{\ed}[1]{
#1
}

\newcommand{\vect}[1]{%
  \mathbf{#1}
}
\newcommand{\matr}[1]{%
  \mathbf{#1}
}
\newcommand{\dimless}[1]{%
  \hat{#1}
}
\newcommand{\Log}[1]{\log\left(#1\right)}

\begin{document}

\title{Loss of control in pattern-directed nucleation: a theoretical study}

\date{\today}

\author{Felix Kalischewski}
\affiliation{\frenchspacing Westf\"alische Wilhelms Universit\"at M\"unster, Institut f\"ur physikalische Chemie, Corrensstr.\ 30, 48149 M\"unster, Germany}
\affiliation{\frenchspacing Center of Nonlinear Science CeNoS, Westf\"alische Wilhelms Universit\"at M\"unster, Germany}
\affiliation{\frenchspacing NRW Graduate School of Chemistry, Westf\"alische Wilhelms Universit\"at M\"unster, Germany}
\author{Jia Zhu}
\affiliation{\frenchspacing Westf\"alische Wilhelms Universit\"at M\"unster, Physikalisches Institut, Germany}
% Wilhelm-Klemm-Str. 9, 48149 M\"unster,
\author{Andreas Heuer}
\affiliation{\frenchspacing Westf\"alische Wilhelms Universit\"at M\"unster, Institut f\"ur physikalische Chemie, Corrensstr.\ 30, 48149 M\"unster, Germany}
\affiliation{\frenchspacing Center of Nonlinear Science CeNoS, Westf\"alische Wilhelms Universit\"at M\"unster, Germany}
\affiliation{\frenchspacing NRW Graduate School of Chemistry, Westf\"alische Wilhelms Universit\"at M\"unster, Germany}

\begin{abstract}
The properties of template-directed nucleation are studied close to the transition where full nucleation control is lost and additional nucleation occurs beyond the pre-patterned regions. First, kinetic Monte Carlo simulations are performed to obtain information on a microscopic level. Here the experimentally relevant cases of 1D stripe patterns and 2D square lattice symmetry are considered. The nucleation properties in the transition region depend in a complex way on the parameters of the system, i.e.\ the flux, the surface diffusion constant, the geometric properties of the pattern and the desorption rate. Second, the properties of the stationary concentration field in the fully controlled case are studied to derive the remaining nucleation probability and thus to characterize the loss of nucleation control. Using the analytically accessible solution of a model system with purely radial symmetry, some of the observed properties can be rationalized. A detailed comparison to the Monte Carlo data is included.
\end{abstract}

\pacs{08154711}


\maketitle

\section{Introduction}
The controlled fabrication of nano-structures on surfaces has achieved a lot of interest during the past decade because of its great application potential in the area of micro-electronics. Predefined structures at this scale can in principle be obtained in a number of ways: by self-organizing growth, by photo-lithography and etching, and by template directed nucleation. The occurrence of self-organizing growth mechanisms is relatively rare and the obtained patterns are by no means arbitrarily controllable; the process of lithography and etching has become very sophisticated and versatile, but unfortunately it is not applicable to every kind of substrate; many organic substances for example cannot be patterned in this way.

This is the point where template directed growth becomes interesting: in this technique the areas of desired accretion are marked in some way to favor the nucleation of the substance, which is subsequently deposited. This way, the patterning is achieved indirectly by means of an already existing technique. The preferential nucleation sites can be created in numerous ways: the flat surface can be roughened at the points of interest by techniques such as focused ion beam engraving \cite{BardottiNanoclusters}, laser ablation \cite{LaserAblation}, direct imprinting \cite{NatureDirectImprinting}, or even by the tip of atomic force microscopes \cite{AFMPatterningHyon,AFMPatterning}; alternatively substances with a comparatively higher binding energy to the depositable substance can be applied in a controlled fashion to provide a pre-defined nucleation site. This can be accomplished e.g.\ by printing \cite{BrisenoNature}, by dip-pen writing \cite{DipPenWritingLehnert}, or by lithographic methods \cite{ChiPRL}.

\begin{figure}[ht]
\includegraphics[width=0.75\linewidth]{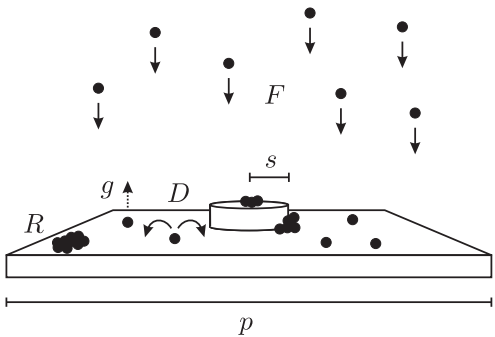}
\caption{\label{FIG:SchemeMechanism}Schematic representation of the considered mechanisms.}
\end{figure}

The specific process analyzed in this work is that of template-directed nucleation as applied by \name{Wang} \etal\ \cite{ChiPRL} However, due to the general nature of the underlying mechanisms, the conclusions drawn in the course of the paper should also be applicable to a wide range of similar techniques. In their experiments a silicon oxide surface was patterned with gold dots electron beam lithography. Then organic substances with a comparably higher affinity to the gold pattern were deposited by molecular beam. Following adsorption these molecules diffused along the surface to eventually either nucleate on a gold dot, to form a new two-dimensional nucleus in combination with other adatoms, or to desorb as visualized in figure~\ref{FIG:SchemeMechanism}. The experiments were carried out at varied pattern spacing and for several organic substances, of which the deposition of diferrocene exhibited the least fluctuations.

From an application point of view one is interested in a maximal range of full nucleation control. Specifically this means: how far can (at constant deposition flux and temperature) the pre-patterned gold sturctures be placed apart until additional nucleation will occur aside of the pattern? This work focuses on the theoretical understanding and description of this transition region, which, as a first approximation, one expects to occur, when the density of the pattern matches the saturated island density that would be obtained from a nucleation experiment on an unpatterned surface under the same conditions \cite{BardottiNanoclusters}. On second thought, however, a sensible and regular array of sinks should provide a better `drainage' of the deposited material and hence an increased nucleation control.

\ed{
We approach this problem by means of two complementary theoretical methods. In section~\ref{HL:MC} we present a detailed description of the MC simulation applied to model the experiment. Then, in section~\ref{HL:NucleationControl} introduce the dimensionless notation from Ref.~\onlinecite{ChiPRL}, which will be applied here, as well, and we discuss the qualitative behavior of the nucleation control.

To obtain a theoretical understanding of the limit of nearly full nucleation control, in section~\ref{HL:Theory} we apply a model similar to the level-set methods of Ref.~\onlinecite{VardavasGrowthWithDefectSites,ChenLevelSetEpitaxial,PhysRevE.58.R6927}. In this approach the deposited adatoms are assumed to diffusive independently such that their concentration can be described in BCF-like \cite{BCF} fashion by a continuous concentration field, to which the pre-patterned sites provide the level-set boundaries. The nucleation probability is then derived on the basis of standard nucleation theory \cite{VenablesPhilosMag,IslandsMoundsAtoms,EvansThielBarteltReview,MaassReview}. Specifically, the local nucleation rate is a function of the adatom concentration and hence the overall nucleation probability can be expressed by a corresponding integral over the surface. On this basis, the nucleation behavior is then discussed in the one- and two-dimensional case analytically as well as numerically. As the deposition flux is experimentally easily accessible, this is done with special respect to the flux dependence.}

\section{Monte Carlo Simulations\label{HL:MC}}

To gain deeper insight on the atomistic mechanisms of pattern directed growth we examine the behavior of patterned systems by means of a Monte-Carlo (MC) simulation. Since we are not interested in the short-time ballistic movement of the adsorbed particles, a MC algorithm is sufficient to model the diffusive behavior\cite{LeviKotrlaCrystalGrowthReview}. It has been shown \cite{HuitemaMCMD,WeinbergMCMD} that a MC step of a diffusive process can be associated to a corresponding mean time step, allowing a dynamical interpretation. In contrast to the alternative of a molecular dynamics (MD) simulation, a MC algorithm provides the additional possibility of enhanced computational efficiency due to exclusion of `spectator-particles' from the propagation routines, \ed{the substrate particles in this case.}

Since the molecular dimensions are much smaller than the relevant environmental length scales we use simple \name{Lennard-Jones} (LJ) pair potentials to mimic the molecular interactions. The parameters $\sigma$ and $\epsilon$ are a measure for the atom-atom distance and depth of the potential well, respectively. The substrate (S), the pattern (A), and the deposited material (M) are taken into account as individual particles. To enhance computational efficiency, the chosen LJ variant possesses a cutoff $\rc$, above which the interaction energy is zero and interactions can be neglected:
\begin{align}
V(r) = 4\,\epsilon \cdot \Bigg[  &\left(\frac{\sigma}{r}\right)^{12} - \left(\frac{\sigma}{r}\right)^6 \nonumber\\
&+ (r-\rc)\left(\frac{12\,\sigma^{12}}{\rc^{13}} - \frac{6\,\sigma^6}{\rc^7}\right) \nonumber\\
&+ \left(\frac{\sigma}{\rc}\right)^6 - \left(\frac{\sigma}{\rc}\right)^{12} \Bigg],\quad \mrm{for} \quad r<\rc
\end{align}
The additional terms ensure a continuous transition of potential and force at $\rc$ while showing only a minor influence on the potential otherwise. 

With all of the A and S particles fixed, the relevant interactions are reduced to the ones involving M particles. The chosen LJ parameters are $\epsilon_\mrm{MA}/\kT=5$, $\epsilon_\mrm{MM}/\kT=3.7$ and $\epsilon_\mrm{MS}/\kT=1.5$, while for the sake of simplicity all radial constants are set to $\sigma=0.890\,a$, with $a$ being the fcc lattice constant and $\rc=2.5\,\sigma$. The energy parameters were chosen such as to prefer the accretion of M on A, as well of M on M over the simple adsorption on the substrate. The M-S interaction is a compromise between a sufficiently high surface diffusion rate on the one hand and at reasonably low desorption rate on the other.

We apply off-lattice displacements. Per sweep every particle is subjected to a trial move, which is a product of a randomly generated vector and an also stochastically determined jump length. The normalized displacement vector is chosen from a uniform distribution on a sphere \cite{Marsaglia}, while the jump length is a uniform deviate in the interval $[0.01\,\sigma,0.5\,\sigma]$. On the cost of computational efficiency, this interval is chosen small enough to prevent frequent `tunneling' of particles through energetic barriers.

\begin{figure}[ht]
\includegraphics[width=\linewidth]{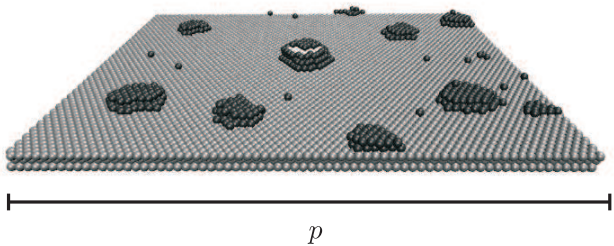}
\caption{\label{FIG:MCDotPic}Typical simulation setup for a pre-patterned surface with dots on a squared lattice. The substrate has the dimensions of $70\times35$ unit-cells. The core of the middle cluster is built from pattern material depicted in white.}
\end{figure}

The simulation is conducted in a simulation box with periodic boundary conditions. A typical simulation setup consists of a substrate located at the middle of the $z$-axis covering the $xy$-plane, yielding a quasi-infinite surface. Since the chosen potential is radially isotropic the possible crystal lattices are fcc or hdp. However, neglecting the proper structure of the material should not pose any problem: firstly, because simple lattices have previously been successfully used to model the growth of different materials \cite{LeviKotrlaCrystalGrowthReview}; and secondly because all necessary physical properties -- especially the preference of the M-A interaction over the M-S alternative -- are included. The lattice of the substrate is oriented with $(111)\,||\,\vect{z}$. The `unit-cells' resulting from this orientation have the dimensions of $a\times \sqrt{3}\,a \times \sqrt{6}\,a$ and contain six particles in three layers. The substrate is then patterned with three more layers of A particles in the desired morphology. As above, the same lattice constant is used for both, S and A particles. A typical setup resembling the experimental conditions of Ref.~\onlinecite{ChiPRL} is depicted in figure~\ref{FIG:MCDotPic}.

\begin{figure}[ht]
\includegraphics[width=\linewidth]{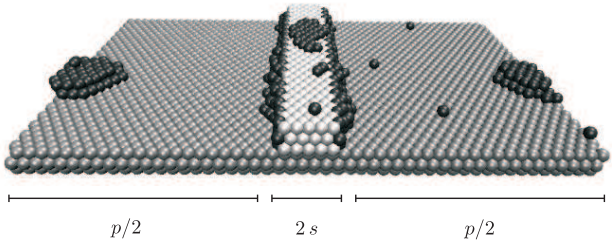}
\caption{\label{FIG:MCStripePic}A typical simulation setup consisting of a substrate with dimensions $50\times20$ as well as a pattern stripe on the  substrate. Close to the middle between the stripes -- with respect to the periodic boundary conditions -- a new nucleus is formed.}
\end{figure}

The deposition flux is modeled by periodic addition of M particles to the substrate. The $x$- and $y$-coordinate of the deposition point are determined uniformly within the dimensions of the simulation box. The particles are then inserted at maximal $z$ and subsequently moved toward the substrate, until an energy minimum is met. After this, the particles are subject to the described MC propagations. Desorption is included \ed{as well}: the system is frequently probed for a particle out of contact ($r>\rc$) with any other particle. If such a particle is found, it is considered desorbed and removed from the simulation.

The simulations, with the exception of the latter ones of section~\ref{HL:2DSimulation}, were conducted at a deposition rate of $F=1.43\cdot10^{-7} \mrm{pt}/(\sigma^2\,\mrm{step})$, where $\mrm{pt}\;\hat{=}\;\mrm{particles}$ and `step' refers to a MC-step. The diffusion coefficient was determined by analysis of the mean square displacement of non-interacting M particles ($\epsilon_\mrm{MM}=0\,\kT$) on an unpatterned surface and found to be $D=3.50\cdot10^{-3}\,\sigma^2/\mrm{step}$. With a resulting rate of $D/F=2.45\cdot10^{4}\sigma^4/\mrm{pt}$, which corresponds to $D/F=1.15\cdot10^4$ in the units of \name{Brune} \cite{BruneReview}, the considered setup lies one order of magnitude below the condition of $D/F>10^{5}$ assumed there for the theoretical treatment of non-interacting particles.
This in addition to the inclusion of desorption which is also typically neglected \cite{VenablesPhilosMag,IslandsMoundsAtoms,EvansThielBarteltReview,MaassReview,BruneReview} can lead to deviations from the ideally expected behavior. As, however, an increase of the temperature would not only raise $D$ but desorption as well, and a sufficient decrease of $F$ would significantly reduce the simulation efficiency we remain with the presented parameters.

The desorption coefficient was determined from a similar simulation: after equilibration adsorption and desorption flux are equal requiring $F=g\,c$. The desorption coefficient $g=12.7\cdot10^{-6}/\mrm{step}$ could hence be derived from the average equilibrium adatom concentration. Consequently, the influence of desorption cannot be neglected with respect to the stationary concentration (for more details see equation~\eqref{EQ:Conc1DWtDeseq}). These parameters and a mean system size of $p=40\,\sigma$ in combination with the considerations of appendix~\ref{HL:APPStationaryConcRegime}, equation~\eqref{EQ:StationaryTime} lead to an estimate of $50\,000$ steps to reach a stationary concentration profile in the 1D-case. This lies well within the typical simulation time of $2-4\cdot10^6$ steps and is sufficiently below $3\cdot10^5$ steps, the time at which the substrate becomes saturated with islands.

\section{Nucleation Control\label{HL:NucleationControl}}
We regard the loss of nucleation control an a quadratic pattern as experimentally observed in Ref.~\onlinecite{ChiPRL} with special respect to the transition region, where control is gradually lost. To this purpose we simulated substrates constructed of $n \times n/2 \times 1$ `unit-cells', where the ratio of $x$- to $y$-dimension is chosen to approximate the desired square lattice by means of the non-square `unit-cells'. A dot of radius $s=3\,a$ and three layers of height built from the same lattice are added to the middle of the surface.

\subsection{Dimensionless units}
To evaluate the obtained data and to allow the direct comparison of different experimental as well as simulated systems, a dimensionless representation in analogy to Ref.~\onlinecite{ChiPRL} is introduced. We define the dimensionless nucleation efficiency as the number of prepatterned islands divided by the number of the actually obtained nuclei: for a single unit-cell of a square pattern the nucleation control is hence given by 
\begin{equation}\label{EQ:DefXn}
\xn := \frac{1}{1+R}
\,,
\end{equation}
where $R$ is a fractional number reflecting the averaged number of additional nuclei per unit cell.

The spatial coordinate is renormalized by the characteristic length scale $\lambda$, which is derived from the nucleus density of an unpatterned surface:
\begin{equation}\label{EQ:DefLambda}
\lambda := \sqrt{\left(\frac{A}{N}\right)_\mrm{u}}
,
\end{equation}
where $A$ is the surface area. Using $\lambda$ the lattice constant $p$ of the pattern can be expressed in the dimensionless form:
\begin{equation}\label{EQ:DefPstar}
\tilde{p} := \frac{p}{\lambda}
.
\end{equation}
It should be noted that for patterns with more than one lattice constant the specific choice of $p$ is in principle arbitrary.

\subsection{Limit behavior}
Mechanistically, one can expect growth to occur by a diffusion-aggregation mechanism as depicted in figure~\ref{FIG:SchemeMechanism}. \ed{After deposition, the adatoms explore the surface diffusively, until they meet or form a nucleus or pattern site, or desorb from the surface altogether.}

On an unpatterned surface nucleation occurs if a sufficient number of particles aggregate to form a stable two-dimensional nucleus, that will then grow by accretion of further particles. On a pre-patterned surface on the other hand, the first nuclei are already supplied by the pattern material, which is chosen for its comparatively more favorable interaction energy with the deposited substance. The subsequent growth process is then an interplay of depletion, diffusion and deposition/desorption. In the limit where $p$ is sufficiently small or $D/F$ accordingly large, all arriving particles are transported to the nuclei and full nucleation control is maintained\cite{VardavasGrowthWithDefectSites,BardottiNanoclusters}. One consequently expects
\begin{equation}\label{EQ:lowerlimit}
\xn = 1 \quad \mrm{for} \quad \tilde{p} \ll 1
.
\end{equation}

Should however the adatom concentration at some point on the surface facilitate the formation of additional nuclei control is gradually lost. This leads to the second limit case: at very large spacing between the dots the existence of the pattern is negligible and the obtained nucleus density approaches the density of an unpatterned system. As the unpatterned system is used for the renormalization of the space coordinate, the expected limit behavior is
\begin{equation}\label{EQ:upperlimit}
\xn = \tilde{p}^{-2} \quad \mrm{for} \quad \tilde{p} \gg 1
.
\end{equation}

\subsection{Transition region}
As a first approximation, one expects the loss of nucleation control to set in, when the density of the patterned surface matches that of its unpatterned equivalent \cite{BardottiNanoclusters}. In the case of the applied quadratic pattern this corresponds to $p=\lambda$ and consequently $\tilde{p}=1$. A sensible array of sinks should however supply a more effective `drainage' of the deposited material, resulting in prolonged nucleation control for values of $\tilde{p}>1$.

\begin{figure}[h!]
%\subfigure[\label{FIG:NewData}.]{
\includegraphics[width=\linewidth]{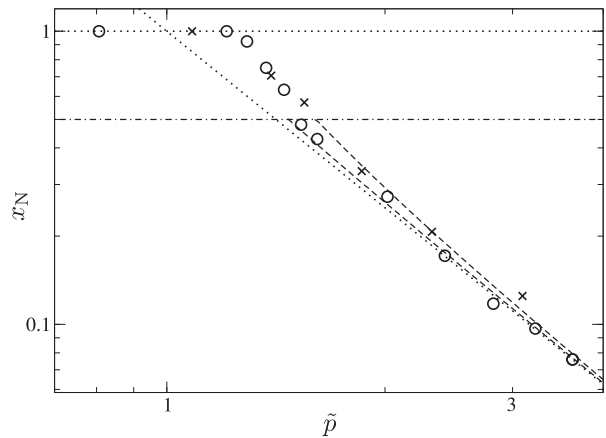}
%}
\caption{{\label{FIG:NewDataXn}}Nucleation control in dependence of the reduced periodicity for the fluxes $1.43\cdot10^{-7}\,\mrm{pt}/(\mrm{step}\,\sigma^2)$ ($\circ$) and $1.43\cdot10^{-6}\,\mrm{pt}/(\mrm{step}\,\sigma^2)$ ($\times$). As before the limit behavior for $\tilde{p}\ll 1$ and $\tilde{p}\gg 1$ is indicated by dotted lines. The dashed-dotted line marks $\xn=1/2$. The dashed lines symbolize the deviation from the limit behavior caused by the dot size according to equation~\eqref{EQ:XnDotCorr} with $\tilde{s}_\mrm{eff}=0.23$ and $0.43$.} 
\end{figure}

To further investigate the behavior at the transition point on the basis of the results in Ref.~\onlinecite{ChiPRL}, we conducted additional simulations at different $p$. The first four systems following the loss of nucleation control were obtained from twelve independent runs while the remaining points were determined from six runs. The number of islands is determined from the stationary regime before the onset of coalescence. The obtained data is depicted in figure~\ref{FIG:NewDataXn} as circles exhibiting an extrapolated loss of nucleation at $\tilde{p}=1.27$ with $\lambda=24.8\,\sigma$.

It can be seen that the first three data points beyond the loss of nucleation control follow in almost linear succession, whereas a jump is observed with respect to the following points, which are closer to the expected limit behavior. The standard deviation of these results, which is about the size of the symbols, suggests that this is not a statistical effect. Closer scrutiny of this jump reveals that it is observed as $\xn$ drops below $1/2$, marking the point where a second additional nucleus is formed and which is indicated by the dashed line. For $\xn>1/2$ only one additional island is formed which is expected to be found close to the corner, the spot of highest concentration and farthest away from the pattern. Any additional nucleation will therefore in principle lead to a new lattice constructed from squares with an edge length smaller by a factor of $1/\sqrt{2}$. This way the area is shared by the islands still quite `effectively'. For $\xn<1/2$ however this behavior breaks down, because the nucleation of the second island discontinues this scheme and $\xn$ approaches the disordered limit behavior. This becomes more so as nucleation control is lost further.

As with growing $\tilde{p}$ the nucleation besides the pattern site becomes more and more statistical, the nucleation control can be approximated by the follwing consideration: taking a dot to occupy an area of $\pi s^2$ in addition to its mean capture area of $\lambda^2$, and assuming the remaining area to be filled in the disordered nucleation limit one obtains
\begin{equation}\label{EQ:XnDotCorr}
\xn = \left( \tilde{p}^2 - \pi\,\tilde{s}^2\right)^{-1}
\,,
\end{equation}
where the reduced radius is defined as $\tilde{s}:=s/\lambda$. However, as $s$ is only of the order of a few particles, the initial accretion of molecules around its boundary introduces a non-negligible size increase. Hence one has to apply an effective radius which can be derived from figure~\ref{FIG:MCDotFig} of section~\ref{HL:2DSimulation}: one obtains $\tilde{s}_\mrm{eff}=0.23$ and $0.43$ which in combination with~\eqref{EQ:XnDotCorr} leads to the dashed lines in figure~\ref{FIG:NewDataXn}. It can be seen that this expression indeed describes the convergence to the limit behavior, which occurs slower with decreasing $\lambda$.

From the experimental point of view, however, one is mainly interested in two questions: (a) how does the point at which nucleation control is lost ($p)$ depend on the deposition flux ($F$)? And (b) how does the mean number of additional nuclei ($R$) change with increasing distance between the predefined dots ($p$)? These questions are approached from a theoretical point of view in the next section.

\section{Theoretical Approach of the Transition Region\label{HL:Theory}}
For the description of the transition region we adopt the following probabilistic picture. We consider a pattern which at given $p,s,F,D$ and $g$ is under (nearly) full nucleation control, i.e.\ $\xn\approx1$. In addition, we neglect the spatial growth of the pattern due to particle accretion and we require the adatom concentration $c(\vect{x})$ to reach a stationary steady state based on the following consideration: according to standard nucleation theory \cite{VenablesPhilosMag,IslandsMoundsAtoms,EvansThielBarteltReview,MaassReview} an unpatterned surface will (after it passed through a regime of rapid nucleation) saturate with respect to the nucleus density ($\dd N/\dd t\approx 0$); hence $c(\vect{x})$ becomes quasi-stationary as well. Because of its good experimental reproducibility and near independence on time we select this regime as reference state. If one now considers a patterned surface of comparable nucleus density, this surface can also be expected to exhibit a stationary adatom concentration. Some basic considerations on the time necessary to reach this state can be found in appendix~\ref{HL:APPStationaryConcRegime}.

However, although the stationary system is supposed to be under full nucleation control, there remains a finite chance for nucleation due to fluctuations in the adatom concentration. Specifically this means that in the considered scenario any additional nucleation is \emph{not} caused by the \emph{dynamic} increase of the concentration field which eventually supercedes a threshold, but only by the remaining small but finite nucleation probability of the \emph{static} concentration field. This probability can be evaluated by methods of standard nucleation theory, according to which the local rate of formation of new nuclei depends (after application of the Walton relation \cite{Walton}) on the concentration as
\begin{equation}
\frac{\dd N(\vect{x})}{\dd t} \propto c(\vect{x})^{\istar+1}
,
\end{equation}
where all clusters of size $\istar$ or less are considered thermodynamically unstable while larger clusters are stable. In analogy to the level set methods of e.g.\ Refs.~\onlinecite{VardavasGrowthWithDefectSites,ChenLevelSetEpitaxial} we determine the rate of nucleation in one unit-cell from the corresponding spatial integral. The mean number of additional nuclei formed up to time $t$ can hence be expressed by temporal integration. Neglecting the time necessary to reach the steady concentration profile we find the direct proportionality
\begin{equation}\label{EQ:DefP}
R \propto t\cdot \int_\mrm{U} \dd A\; c(\vect{x})^{\istar+1} \propto \frac{\Theta}{F} \cdot \int_\mrm{U} \dd A\; c(\vect{x})^{\istar+1}
\end{equation}
with $\mrm{U}$ as the unit-cell and a proportionality constant independent of $p, s, F, D, g$ and $t$. The latter part holds for states of identical coverage $\Theta$.

As the limit of full nucleation control with $\xn=1$ (or $R=0$) is not well defined in this probabilistic approach, we consider a critical $R$ with $0<R\ll1$, which serves to describe the point of transition. One can now examine the dependency of $p, s, F, D,$ and $g$  at constant $R$: if for example the deposition flux $F$ is increased slightly, how will the corresponding $p$ have to change? It should be noted that in the limit $\istar\rightarrow\infty$ only the point of maximal concentration contributes significantly to $R$, which in turn leads to the same result as a deterministic critical concentration approach, where new nuclei are considered to be formed when a concentration threshold is superceded.

The qualitative flux dependency is intuitively accessible: An increase in $F$ results in a decrease of $p$ and vice versa, but the behavior of $\tilde{p}$ also depends on the density of the unpatterned substrate via $\lambda$. Thus, if $p$ and $\lambda$ exhibit different dependencies on $F/D$, or if the behavior of the pattern is influenced strongly by the dot diameter, it is conceivable that the specific point at which nucleation control is lost can be shifted. This behavior is examined in the following section.

\subsection{Nucleus densities\label{HL:AdatomConcentration}}
To predict the point at which nucleation control breaks down we have to find expressions for the nucleus density on unpatterned and patterned surfaces as a function of the growth parameters.

For two-dimensional nucleation on an \emph{unpatterned substrate} standard nucleation theory \cite{IslandsMoundsAtoms} supplies a general model, which is based on rate equations and leads to one of the central relations, namely
\begin{equation}\label{EQ:UnpatternedScaling}
\lambda \propto \left(\frac{D}{F}\right)^{\chi/2} \quad \mrm{with} \quad 1/3\le\chi<1
\end{equation}
at constant coverage $\Theta$. This in turn supplies the desired $F$ and $D$ dependence of $\lambda$, while the exponent is a function of the critical nucleus size $\istar$ according to
\begin{equation}\label{EQ:DefChi}
\chi = \frac{\istar}{\istar+2}
\,.
\end{equation}

In the next step we investigate the nucleation behavior of \emph{patterned surfaces}. Similar to BCF-theory \cite{BCF} we apply the following continuum approach: neglecting the stabilizing effects of unstable clusters we seek the stationary adatom surface concentration satisfying the diffusion equation
\begin{equation}\label{EQ:DiffusionEquation}
\frac{\partial c(\vect{x})}{\partial t} = D \nabla^2 c(\vect{x}) - g\,c(\vect{x}) + F = 0
\,.
\end{equation}
On the basis of this stationary concentration field one can then obtain the mean number of additional nuclei according to equation \eqref{EQ:DefP}. The dots are modeled by localized ideal sinks supplying $c|_\mrm{sink}=0$ as boundary conditions.

 The general shape of the concentration field can be discussed by means of a reduced representation. Introducing dimensionless scales for length and concentration
\begin{equation}\label{EQ:DefDimless}
\vect{\dimless{x}} := \vect{x}/p \quad\mrm{and}\quad \dimless{c} := \frac{2\,D}{p^2F}\,c
\,,
\end{equation}
as well as the dimensionless parameter
\begin{equation}\label{EQ:DefBeta}
\beta:=\sqrt{g/D}\,p
\,,
\end{equation}
equation~\eqref{EQ:DiffusionEquation} can be rewritten as
\begin{equation}
\dimless{\nabla}^2 \dimless{c}(\vect{\dimless{x}}) - \beta^2 \dimless{c}(\vect{\dimless{x}}) = -2
\,,
\end{equation}
supplemented by the corresponding boundary conditions. One can now easily see that any solution fulfilling equation~\eqref{EQ:DiffusionEquation} has the general form of
\begin{equation}\label{EQ:FScalingA}
c(\vect{\dimless{x}}) = \frac{F}{g}\,\frac{\beta^2}{2} \,\dimless{c}(\vect{\dimless{x}},\beta)
\,.
\end{equation}
Thus, for $g\neq0$ the general shape of the concentration field, i.e.\ $\dimless{c}(\dimless{x})$, is completely determined by $\beta(g,D,p)$ in combination with the boundary conditions, while for $g=0$ it depends only on the boundaries. The absolute concentration, however, is always directly proportional to $F$.

We now turn to the desired relation between $p$ and $F$: demanding constant $R$ as given in~\eqref{EQ:DefP} and substituting the scaling behavior of~\eqref{EQ:FScalingA} one obtains
\begin{equation}\label{EQ:R2}
R \propto F^{\istar} \cdot p^{2(\istar+1)+d} \cdot \int_\mrm{\dimless{U}}\dd \dimless{A}\,\dimless{c}(\vect{\dimless{x}},\beta)^{\istar+1}
\,,
\end{equation}
where $d=1,2$ is the dimensionality of the considered system. Since in general the above integral has a complex dependence on $p$ via $\beta$, equation~\eqref{EQ:R2} cannot be written in the form $p = h(F)$. Nevertheless important information can be gained from the logarithmic derivative in analogy to equation~\eqref{EQ:UnpatternedScaling}. We define
%\begin{subequations}
%\begin{align}
%a  &:= \left(\frac{\partial \Log{F}}{\partial \Log{p}}\right)_R \label{EQ:Defa}\\
%   %&\;=  -2\frac{\istar+1}{\istar} - \frac{1}{\istar} \left[d + \frac{\partial}{\partial \log p} \log\int_\mrm{\dimless{U}}\dd \dimless{A}\;\dimless{c}(\vect{\dimless{x}},\beta)^{\istar+1} \right]\\%\label{EQ:DefaExplicit}
%   &\;=  \frac{-1}{\istar}\left[ 2\,(\istar+1) + d + \left(\frac{\partial}{\partial \log p} \log\int_\mrm{\dimless{U}}\dd \dimless{A}\;\dimless{c}(\vect{\dimless{x}},\beta)^{\istar+1}\right)_R \right]\label{EQ:DefaExplicit}
%,
%\end{align}
%\end{subequations}
\begin{align}
a  &:= \left(\frac{\partial \Log{F}}{\partial \Log{p}}\right)_R \label{EQ:Defa}\\
\begin{split}
   &\;=  -\frac{1}{\istar}\,\Bigg[\, 2\,(\istar+1) + d \\
   &\quad \quad \quad \quad  + \left(\frac{\partial}{\partial \log p} \log\int_\mrm{\dimless{U}}\dd \dimless{A}\;\dimless{c}(\vect{\dimless{x}},\beta)^{\istar+1}\right)_R \Bigg]\label{EQ:DefaExplicit},
\end{split}
\end{align}

with the help of which the behavior of the transition point can be expressed as
\begin{equation}\label{EQ:Dofa}
\tilde{p} \propto F^{\chi/2+1/a}
\,.
\end{equation}
As the important information is contained in the exponent we additionally introduce
\begin{equation}\label{EQ:Defb}
b := \frac{\chi}{2}+\frac{1}{a}
\,,
\end{equation}
\ed{which is discussed in detail by means of numerical as well as analytical methods in section~\ref{HL:TDAnalyticalAndNumerical}.}

Analogously, the relation between $R$ and $p$ at constant flux is obtained from
\begin{subequations}
\begin{align}\label{EQ:Deff}
f &:= \left(\frac{\partial \Log{R}}{\partial \Log{p}}\right)_F\\ 
  &\;= -\istar\,a \label{EQ:DeffSecond}
\,.
\end{align}
\end{subequations}
\ed{This relation provides a useful means of comparison to the simulation data discussed in section~\ref{HL:2DSimulation} with special respect to figure~\ref{FIG:NewDataR}.}

\subsection{One-dimensional pattern}
\subsubsection{Analytical results}
The one-dimensional case can be interpreted as a pattern of parallel stripes with spacing $p$ between them and a width of $2\,s$ as depicted in figure~\ref{FIG:MCStripePic}. In the dimensionless notation of equation~\eqref{EQ:DefDimless} the value of $s$ is of no relevance, however, as the boundary conditions are supplied by the sinks with $\dimless{c}(\dimless{x}=0)=\dimless{c}(\dimless{x}=1)=0$. For $g=0$ the stable concentration profile is given by the simple parabolic equation
\begin{equation}\label{EQ:Conc1Deq}
\dimless{c}^\mrm{1D}(\dimless{x}) = \dimless{x}-\dimless{x}^2
\,,
\end{equation}
which possesses a maximum of
\begin{equation}\label{EQ:Conc1DMax}
c_\mrm{max}^\mrm{1D} = \frac{p^2 F}{8\,D}
\end{equation}
at $x=p/2$.

Including desorption with $g>0$ identical boundary conditions lead to the stationary solution
\begin{equation}\label{EQ:Conc1DWtDeseq}
\dimless{c}^\mrm{1D} = \frac{2}{\beta^2} \left( -\gamma \exp\left[\beta\, \dimless{x}\right] -\left(1 - \gamma \right)\exp\left[-\beta\, \dimless{x}\right] +1 \right)
\,,
\end{equation}
with $\beta$ as defined in~\eqref{EQ:DefBeta} and $\gamma:=(1+\exp[\beta])^{-1}$. The maximum is found to be
\begin{equation}\label{EQ:Conc1DWtDeseqMax}
c(\dimless{x}=1/2) = \frac{F}{g} \left( 1 -2\,\frac{\Exp{\beta/2}}{1+\Exp{\beta}}\right)
\,.
\end{equation}
In the limit $g\rightarrow 0$ equations~\eqref{EQ:Conc1DWtDeseq} and \eqref{EQ:Conc1DWtDeseqMax} correspond to~\eqref{EQ:Conc1Deq} and~\eqref{EQ:Conc1DMax}, respectively, as can easily be seen from a Taylor expansion at $p/2$. With growing $\beta$ higher order terms begin to contribute, but below $\beta \sim 1$ their influence is rather small and the parabolic shape is retained in general. At further increase of $\beta$ the function starts to develop a plateau between the sinks. The maximal achievable concentration in the case $\beta\rightarrow\infty$ corresponds to an equilibrium in the absence of sinks given by $F/g$ or correspondingly $2/\beta^2$ in dimensionless scales. 

Here, we refrain from an evaluation of equation~\eqref{EQ:Defa} for the 1D case, which can be easily done analytically for $\beta=0$ or numerically for $\beta\neq 0$. Rather, to get a better understanding of the stationary approximation, we analyze the concentration fields of the simulation.

\subsubsection{\label{HL:MCStriped}Simulation}
We simulated surfaces with a striped pattern to evaluate the validity of the mathematical considerations from section~\ref{HL:AdatomConcentration}. The simulated systems consist of a substrate built from $(30,40,50) \times 20 \times 1$ unit-cells. The pattern stripe is constructed from the same lattice; it covers the $y$-dimension and possesses a width of $6\,a$ along $x$. As can be seen in figure~\ref{FIG:MCStripePic}, the M particles preferably aggregate at or on top of the step. However, if $p^2 F/D$ is sufficiently high, additional nucleation, usually close to the middle between the stripes, can be observed as expected.

\begin{figure}[!ht]%
%\subfigure[\label{FIG:MCStripeFig30}$p=30\,a$]{
\includegraphics[width=\linewidth]{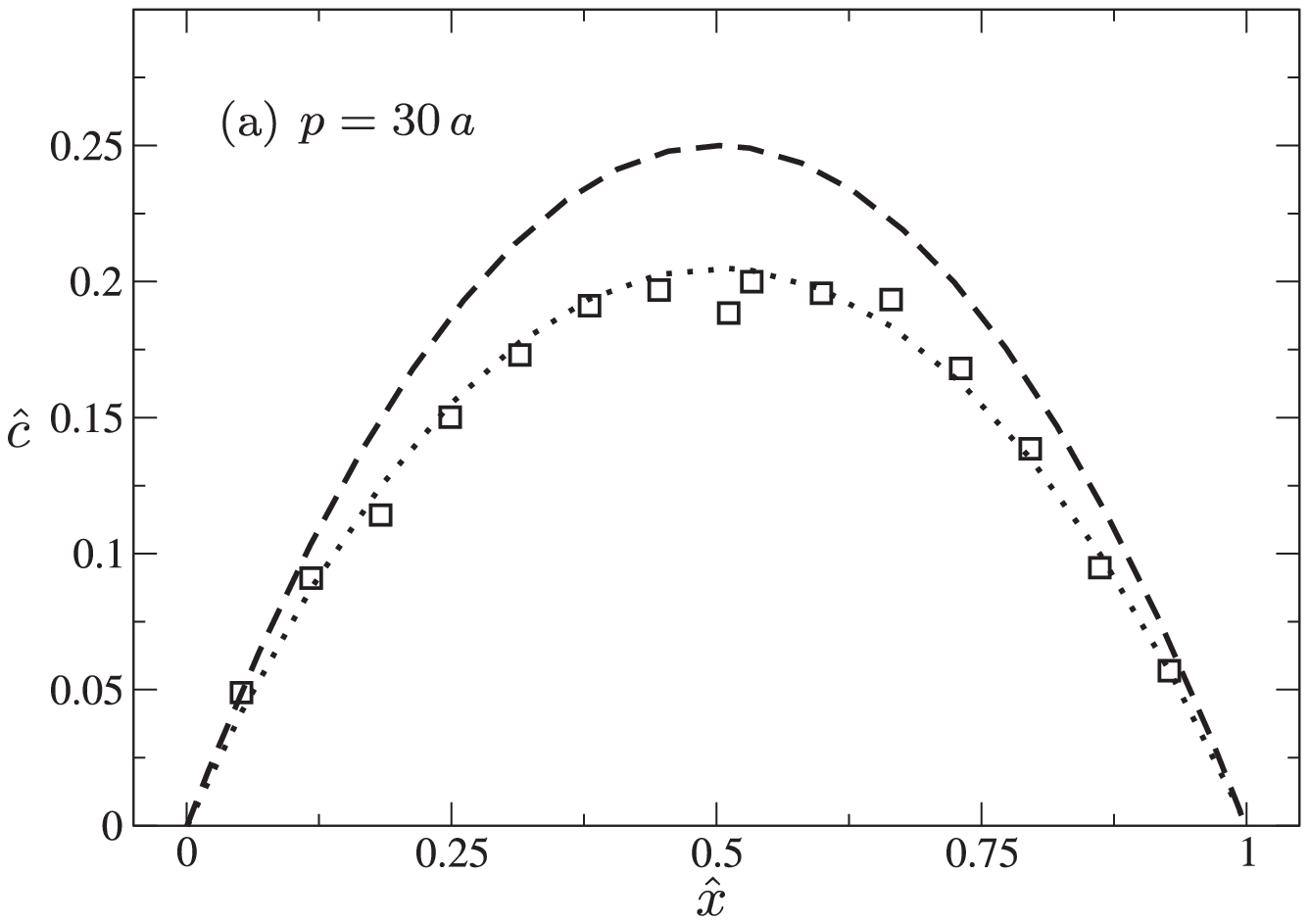}
%}
%\subfigure[\label{FIG:MCStripeFig40and50}$p=40\,a$ ($\circ$) and $50\,a$ ($\times$)]{
\includegraphics[width=\linewidth]{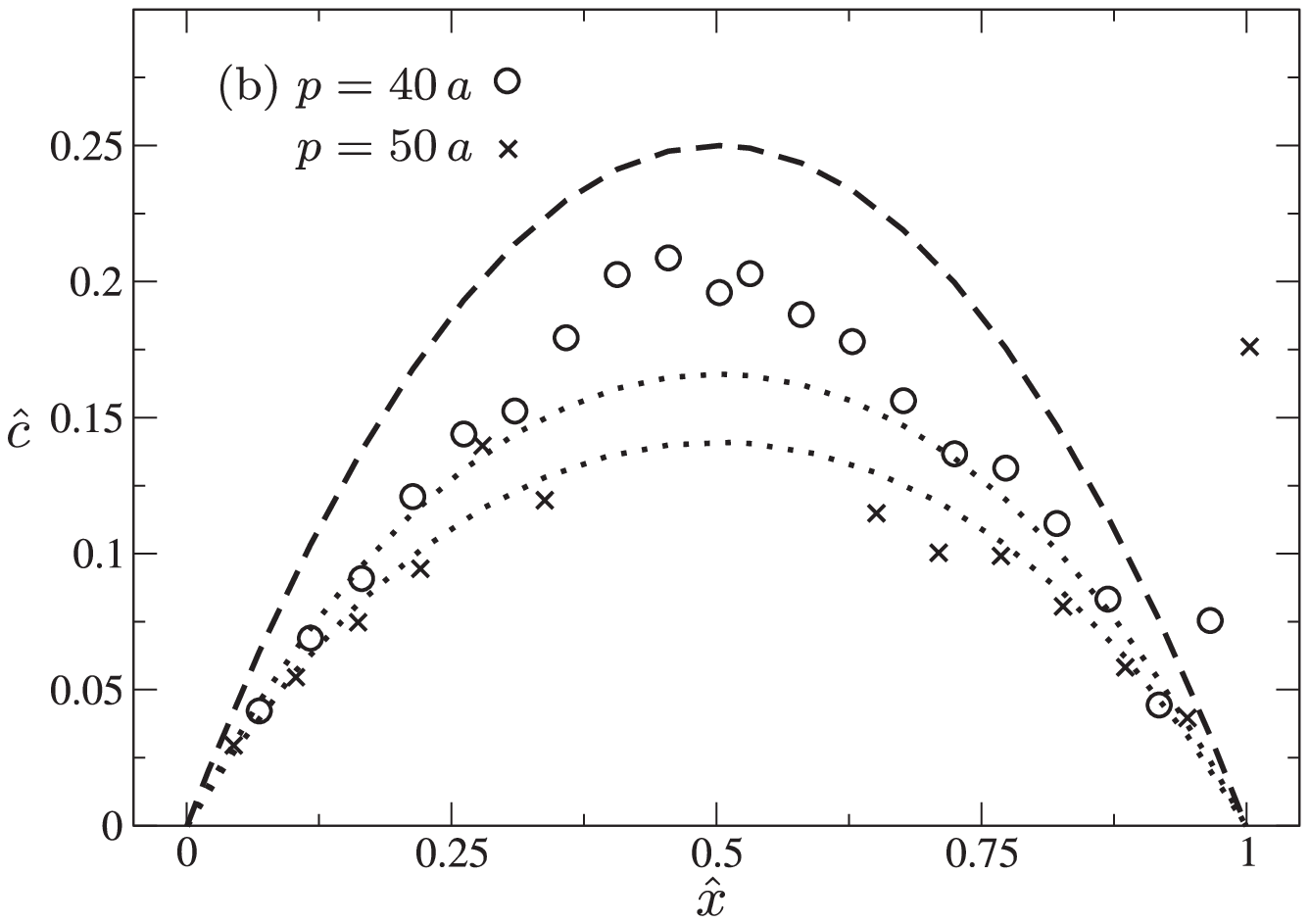}
%}
\caption{\label{FIG:MCStripeFig}Mean reduced concentration for different distances between pattern stripes. The dashed line represents the ideal behavior without desorption according to equation~\eqref{EQ:Conc1Deq}. The dotted lines include desorption according to equation~\eqref{EQ:Conc1DWtDeseq}. For $p=30\,a$ the results are close to the expected behavior. At $p=40\,a$ stabilization of adatoms causes a domain of higher concentration close to the middle. At $p=50\,a$ nucleation sets in resulting in a peak around $\dimless{x}=0.5$ where $\dimless{c}\gg0.25$.}
\end{figure}

The mean adatom concentration along the $x$-axis was determined by an average over the full length of the simulation time and $y$-axis. The results are given in figure~\ref{FIG:MCStripeFig}. At small distances between the stripes ($p=30\,a$) the obtained concentration dependence exhibits the predicted behavior of non-interacting adatoms as given by equation~\eqref{EQ:Conc1DWtDeseq} (dotted line). When the steps are placed farther apart two effects can be observed: (a) the influence of desorption on the concentration profile becomes more pronounced, which can be seen in comparison to the desorption-less concentration profile from equation \eqref{EQ:Conc1Deq} (dashed lines); and (b) mutual stabilization of the adatoms becomes more important. At $p=40\,a$ localized regions of higher concentration due to the formation of unstable clusters appear. This stands in good agreement with the visual observations frequently showing clusters of two or sometimes three atoms in the vicinity of the point of highest concentration. Finally, at $p=50\,a$ nucleation sets in leading to a steep increase in concentration where $\dimless{c}\gg0.25$. At this point it should again be noted that the averaged concentration profiles do not represent a single configuration but represent the mean of the system before and after nucleation. Additionally it should be kept in mind that only the reduced concentrations decrease with increasing $p$; the absolute values behave oppositely.

\subsection{Two-dimensional pattern}
\subsubsection{\label{HL:TDAnalyticalAndNumerical}Analytical and numerical results}
In the one-dimensional case the concentration at the sink has to be zero and consequently the step width $s$ does not have any influence on the concentration fields. In two dimensions however, the relation of sink size to pattern spacing is of importance. With the radial coordinate $r$ and dot radius $s$ as system parameters we again use dimensionless scales:
\begin{equation}
\dimless{r} := r/p \quad \mathrm{and} \quad \dimless{s} := s/p
\end{equation}
Unfortunately, the analytic evaluation of $c(\vect{x})$ is only available for a few limit cases. We therefore first adopt the approximation of a radially isotropic concentration field, leading to the boundary conditions
\begin{equation}
c(\dimless{r}=\dimless{s})=0 \quad \mrm{and} \quad \left.\frac{\dd c(\dimless{r})}{\dd \dimless{r}}\right|_{\dimless{r}=\frac{1}{2}}\hspace{-2ex}=0
\,.
\end{equation}
In polar coordinates and without desorption equation~\eqref{EQ:DiffusionEquation} becomes
\begin{equation}
\frac{1}{r} \frac{\dd}{\dd r}\left(r\frac{\dd c}{\dd r}\right) = -\frac{F}{D}
\,,
\end{equation}
which after twofold integration yields
\begin{equation}\label{EQ:cisoeqn}
\dimless{c}^\mrm{iso}(\dimless{r},\dimless{s}) = \frac{1}{2}\Big( \dimless{s}^2 - \dimless{r}^2 \Big) + \frac{1}{4}\ln\left(\frac{\dimless{r}}{\dimless{s}}\right)
\end{equation}
with the maximum
\begin{equation}\label{EQ:cmaxeqn}
\dimless{c}_\mrm{max}^\mrm{iso} = \frac{1}{4} \left( -\ln\left(2\dimless{s}\right) + 2\dimless{s}^2 - \frac{1}{2} \right)
\end{equation}
at $\dimless{r}=1/2$.

On the basis of this ansatz one can discuss the mathematically simple, but experimentally impractical desorptionless scenario where $s$ and $p$ are scaled varied proportionally, such that $\dimless{s}=\mrm{const}$: in this case the general shape of the stationary reduced concentration field remains unchanged as can be seen from equation~\eqref{EQ:FScalingA}; thus in combination with~\eqref{EQ:DefaExplicit} and~\eqref{EQ:Defb} one obtains
\begin{equation}
a=-2\,\frac{\istar+2}{\istar} \quad \text{and} \quad b=0
\,.
\end{equation}

\begin{figure*}[t]
\newlength{\fxfigwidth}
\setlength{\fxfigwidth}{0.45\linewidth}
%\subfigure[\ $\beta=0$]{\label{FIG:DNumericalWtbeta0}
\includegraphics[width=\fxfigwidth]{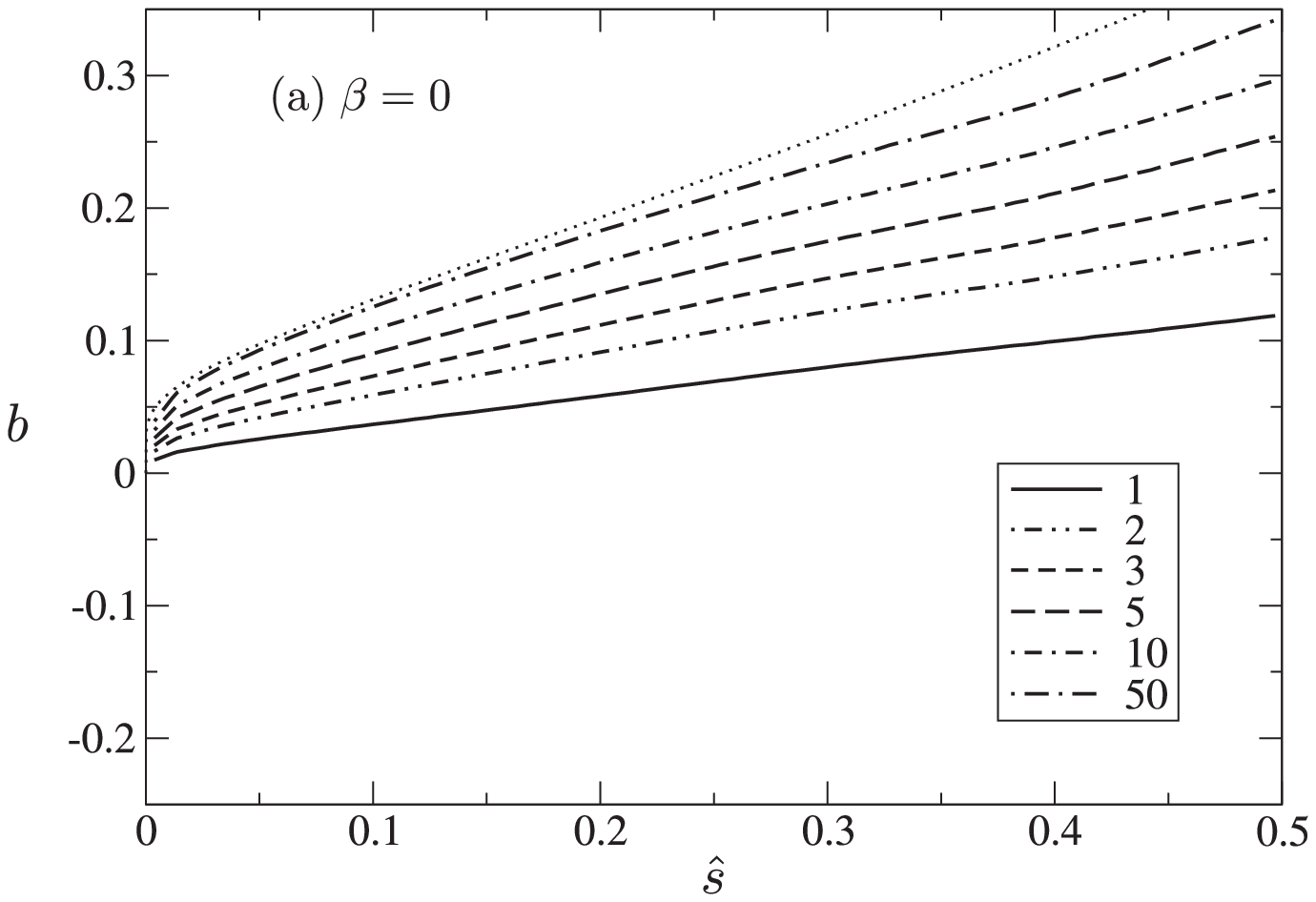}
%}
%\subfigure[\label{FIG:DNumericalWtbeta1.90}\ $\beta=1.90$]{
\includegraphics[width=\fxfigwidth]{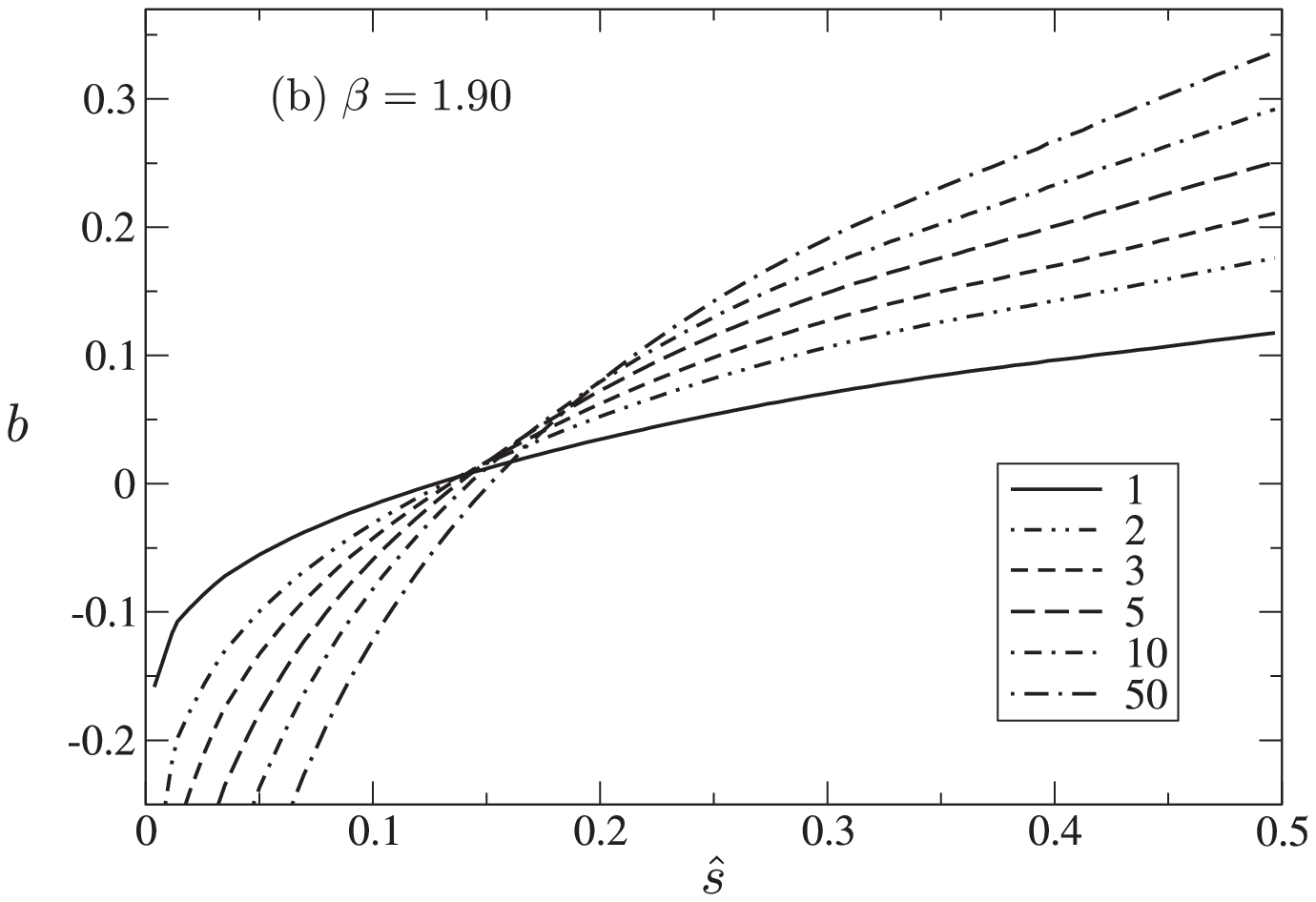}
%}
\par
\vspace*{2ex}
%\subfigure[\label{FIG:DNumericalWtbeta5}$\ \beta=5$]{
\includegraphics[width=\fxfigwidth]{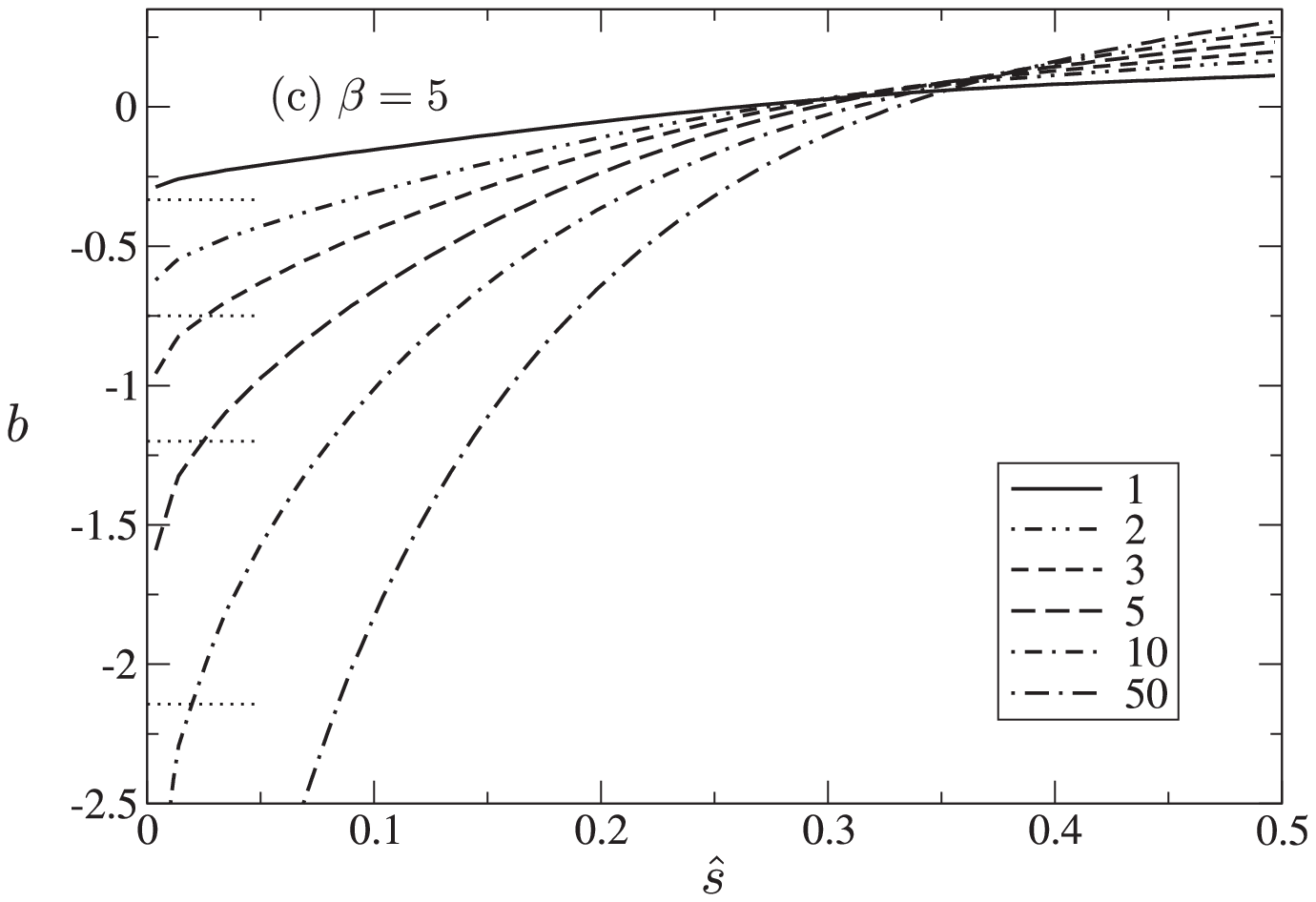}
%}
%\subfigure[\label{FIG:DNumericalWtbeta50}\ $\beta=50$]{
\includegraphics[width=\fxfigwidth]{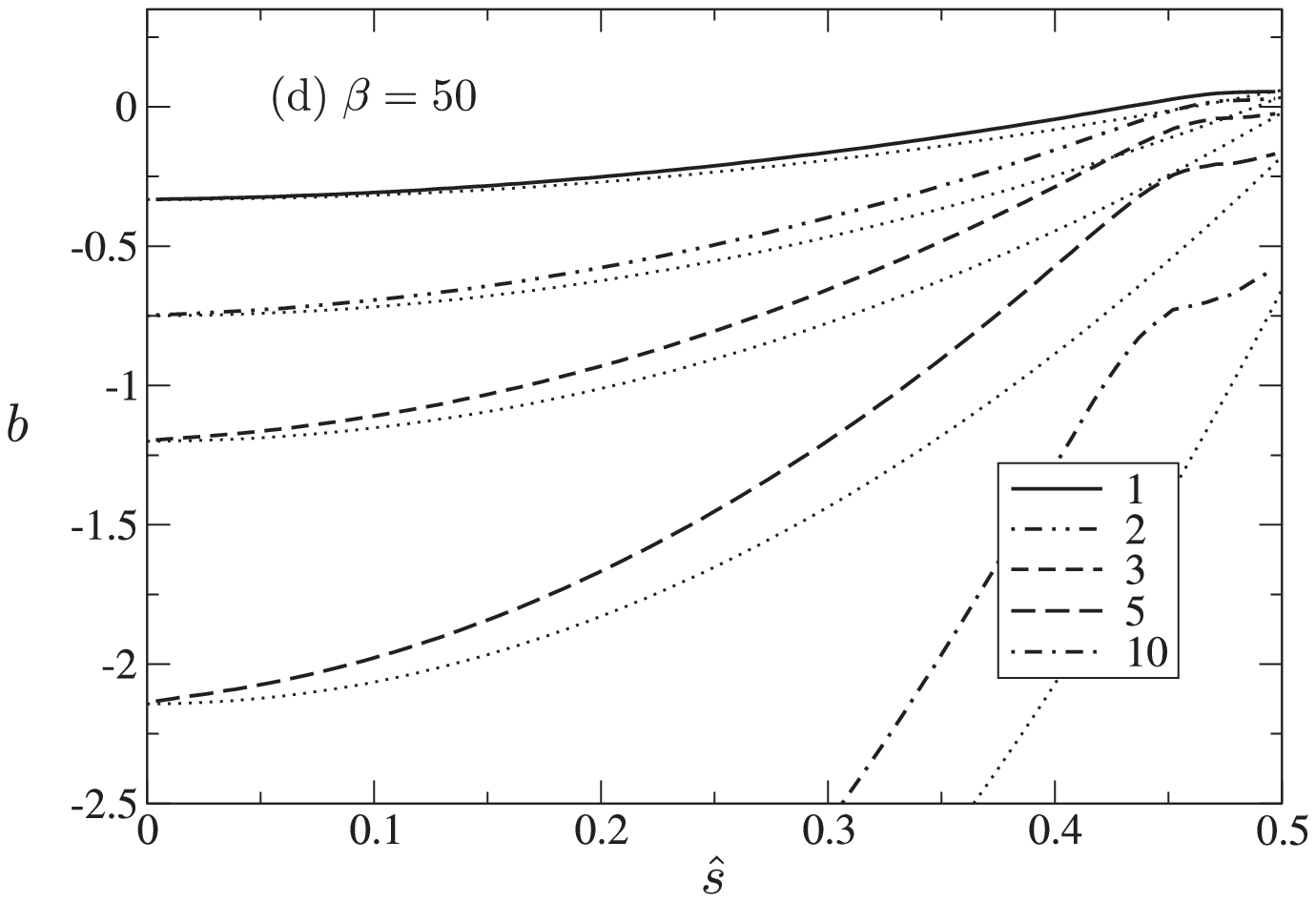}
%}
\caption{\label{FIG:DNumerical}Numerical results for $b(\dimless{s})$ at different $\beta$ as well as different critical nucleus sizes ($\istar$), which are indicated by the legends. The thin dotted lines indicate the expected limit behavior according to equation~\eqref{EQ:bNoDesInfI} for (a) and equation~\eqref{EQ:bForLargeBeta} for (c) and (d), where for (c) only the limit $b(0)$ is depicted.}
\end{figure*}

The experimentally more realistic case where $p$ is changed at constant absolute dot size $s$ is analytically accessible in the following limits: first, in case $\istar\rightarrow\infty$ only the point of maximal concentration contributes to $R$. One can thus apply the $c_\mrm{max}$ ansatz using the no-desorption approximation equation~\eqref{EQ:cmaxeqn}. As shown in appendix~\ref{HL:APPStationaryConcNumerics}, this expression can be fitted to successfully describe the numerically observed maximal concentrations. Combination of this expression with~\eqref{EQ:DefP} and subsequent differentiation yields
\begin{equation}\label{EQ:bNoDesInfI}
b = \frac{-1+4\dimless{s}^2}{2\left(\Ln{2}+\Ln{\dimless{s}}\right)}
\,.
\end{equation}
This expression is included into figure~\ref{FIG:DNumerical}(a) as a thin dotted line. For $s\rightarrow0$ one finds $b\rightarrow0$; as $\dimless{s}$ increases the deviations from the assumed radial isotropy become dominant and the approximation loses its validity.

The second analytically accessible setup is found in the limit $\beta\rightarrow\infty$: In this case the influence of the sinks is diminishingly small and every part of the surface that is covered by pattern material has an adatom concentration of $F/g$. The integrated nucleation probability is then given by
\begin{equation}\label{EQ:AnsatzLargeBeta}
R \propto \left(\frac{F}{g}\right)^{\istar}\negmedspace\left(p^2 - \pi\,s^2\right)
\end{equation}
and differentiation according to~\eqref{EQ:Defa} and subsequent substitution~\eqref{EQ:Defb} leads to
\begin{equation}\label{EQ:bForLargeBeta}
b = \frac{1}{2}\left( \frac{\istar}{\istar+2}-\istar\left(1-\pi\dimless{s}^2\right)\right)
\,.
\end{equation}
This relation is depicted in figure~\ref{FIG:DNumerical}(d) for the corresponding $\istar$ by the thin dotted lines.

Alternatively the stationary concentration for arbitrary setups of lattice and dot shape is readily accessible by means of numerical solution: expanding the concentration field to a discrete grid and expressing equation~\eqref{EQ:DiffusionEquation} by explicit finite differences, one obtains a system of linear equations of the form
\begin{equation}\label{EQ:FiniteDifferences}
\left( \matr{D}-g\,\matr{1}\right)\,\vect{c}_\mrm{st} = -\vect{f}
\,,
\end{equation}
where $\matr{D}$ and $\vect{f}$ describe the concentration change with the next time interval due to diffusion and deposition flux, respectively. This set of linear equations can then be solved by standard matrix algebra \cite{Lapack}. On the basis of these solutions one can now determine the scaling behavior of the nucleation in the following way: first, the patterned exponent $a$ is determined by numerical differentiation according to~\eqref{EQ:Defa} at constant $s$, $D$ and $g$ by variation of $p$ by the width of one grid point. However, the overall curves are calculated at constant $\beta$: when at constant $s$ the distance $p$ is increased to realize a different $\dimless{s}$, $g$ is adapted appropriately to obtain the same $\beta$. Second, $b$ is then calculated from $a$ based on equation~\eqref{EQ:Defb}, with $\chi$ depending on $\istar$ as given in~\eqref{EQ:DefChi}. With respect to the following discussion we assume here that desorption caused deviations of $\chi$ from equation~\eqref{EQ:DefChi} can be neglected, which in the framework of the standard nucleation theory is valid in the long-time limit. In any case, more refined estimates of $\chi$ can easily be implemented.

As has been shown in equation~\eqref{EQ:DefaExplicit} $a$ does not depend directly on $p$, but only on $\beta$ and the boundary conditions ($\dimless{s}$). Thus the parameters for the numerical computations can be chosen arbitrarily. Using $\mrm{u}$ and $\mrm{t}$ as the respective arbitrary units of length and time, the chosen parameters a all set to unity: $s=1~\mrm{u}$, $D=1~\mrm{u}^2/\mrm{t}$ and $F=1\,\mrm{pt}/(\mrm{u}^2\mrm{t})$. The diffusion field is discretized on a $145\times145$ grid for every considered $p$. To calculate $a$ the grid is enlarged by one field in each dimension. To evaluate the influence of desorption numerical solutions were found for $\beta=0$ modeling the no-desorption case, $\beta=1.90$ which corresponds to the parameters chosen for the MC simulation presented in section~\ref{HL:MCStriped}, and finally with $\beta=10$ and $50$ two cases of medium high and very high desorption are discussed. The nucleation probability is determined by integration over the stationary concentration profiles with exponents $\istar=1,2,3,5,10$ and $50$ according to equation~\eqref{EQ:DefP}, where $50$ is already close to the $\istar\rightarrow\infty$ limit and thus the $c_\mrm{max}$ ansatz. The obtained data is depicted in figure~\ref{FIG:DNumerical}.

Recapitulating the concepts that lead to figure~\ref{FIG:DNumerical} it can be read in the following way: if one knows a system defined by $R,p,F,D,g,s,\lambda$ and $\istar$, one can use the corresponding graph to determine by means of $b$ how much $\tilde{p}$ will change with an infinitesimal change in $F$ at otherwise constant parameters (especially $R$). For $b>0$ an increase in $F$ leads to a longer retained nucleation control, while the opposite holds for $b<0$.

All of the graphs in figure~\ref{FIG:DNumerical} exhibit a similar tendency of monotonous increase of $b$ with $\dimless{s}$. The desorptionless case is depicted in figure~\ref{FIG:DNumerical}(a). As can be seen, the curves for $\istar=1,3$ and $5$ intersect the abscissa meaning the direction of the scaling behavior can generally change. In the two-dimensional case as $s\rightarrow 0$ the influence of the sinks on the concentration field disappears and the concentration diverges, which can be seen from the numerical treatment in the appendix. This does not require a divergence of $b$ however. It can also be seen that with increasing $\istar$ the curves approach the analytically approximated behavior for $\istar\rightarrow\infty$ from equation~\eqref{EQ:bNoDesInfI}, which is included as the thin dotted line. To yield an appropriate approximation the corresponding fit parameter from table~\ref{TAB:hfitpars} in appendix~\ref{HL:APPStationaryConcNumerics} is used which scales $\dimless{r}\rightarrow A_0\dimless{r}$. It has to be kept in mind however, that the approximation of a radially isotropic concentration field leading to equation~\eqref{EQ:bNoDesInfI} breaks down at large $\dimless{r}$.

In figure~\ref{FIG:DNumerical}(b) one can find the results for $\beta=1.90$, which correspond to the parameters of the MC simulations. The obtained data exhibit a significant qualitative change in the shape of the curves: at large $\dimless{s}$ the distance between the dots becomes smaller which in turn renders the `effective' $\beta$ small; thus the influence of desorption becomes less important and the results are very similar to the ones obtained in the no-desorption case. However, at small $\dimless{s}$ the existence of desorption leads to a distinctly different behavior: as before, with $\dimless{s} \rightarrow 0$ the influence of the sinks disappears, but now instead of diverging the concentration approaches $F/g$.  Consequently the ansatz of equation~\eqref{EQ:AnsatzLargeBeta} holds and the limit for $b(\dimless{s}=0)$ is given by equation~\eqref{EQ:bForLargeBeta}. In this limit the order of the different $\istar$ with respect to $b$ is flipped and the curves intersect. The final convergence is not shown in the graphs however, as it occurs too close to $\dimless{s}=0$.

The approach of the limit at small $\dimless{s}$ becomes more recognizable when $\beta$ is raised to $5$ as depicted in figure~\ref{FIG:DNumerical}(c). Here it can be clearly seen that with $\dimless{s}$ going to zero the obtained data approaches the value of equation~\eqref{EQ:bForLargeBeta} for $\dimless{s}=0$, which is indicated by the thin dotted lines. Considering the other limit $\beta$ is now sufficiently large to show an effect even if the effective distance between the dots decreases. As this effect also moves the curves toward the limit behavior shown by the thin dotted lines in figure~\ref{FIG:DNumerical}(d) the intersection is shifted toward $\dimless{s}=1/2$. If $\beta$ is further increased to $50$ (see figure~\ref{FIG:DNumerical}(d)) the numerical data further approach this limit to the point that the intersections disappear. The deviations in the region of $\dimless{s}\approx0.45$ might be attributed to the fact that close to coalescence of the dots the area with a low effective distance and thus reduced influence of desorption increases. For $\dimless{s}\rightarrow1/2$ the curves seem to approach the theoretical prediction again. However, in this region numerical artefacts caused by the discretization of the comparably small remaining area have to be taken into account, also.

\subsubsection{Simulation\label{HL:2DSimulation}}
\begin{figure}[!ht]
\includegraphics[width=\linewidth]{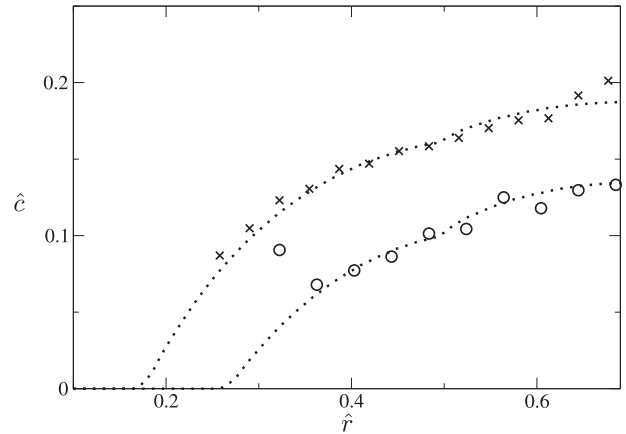}
\caption{{\label{FIG:MCDotFig}}Reduced radial concentration for dot patterns with lattice spacing $n=20$ ($\circ$) and $30$ ($\times$). The dotted lines represent the ideal behavior as calculated by numerical solution. Due to the different sizes the reduced dot radii are $\dimless{s}_\mrm{eff}=0.27$ and $0.18$, respectively.}
\end{figure}

Again, to evaluate the theoretical considerations we examine the concentration profile around the center of the gold dot. To this purpose we simulated two systems with $n=20,30$. The obtained radially averaged concentrations are given in figure~\ref{FIG:MCDotFig} in the dimensionless representation. In contrast to the one-dimensional case we now consider the radially averaged concentration profiles. The average is taken with respect to only one tile of the pattern, meaning from $0\le\dimless{r}\le1/2$ the determination is straightforward, but for $1/2<\dimless{r}\le1/\sqrt{2}$ only the corners of the same square contribute. The dotted lines represent results obtained by numerical solution under the given parameters as described in the previous section. An approximation formula for this behavior is discussed in detail in appendix~\ref{HL:APPStationaryConcNumerics}. \ed{Of course, the depicted results do not represent a single configuration but are the result of an average over the whole simulation.} This leads to some error with respect to the fact that the dot size is not constant, but increases with time \ed{via} continued accretion of new particles. The effective average radius chosen for the numerical solution of both problems is $s_\mrm{eff}=5.65\,\sigma$. With $p$ chosen as the geometric mean of the substrate dimensions of the MC simulation this leads to the reduced radii $\dimless{s}_\mrm{eff}=0.27$ and $0.18$ for $n=20,30$, respectively.

As in the one-dimensional case, the predictions based on non-interacting adatoms stand in good agreement with the simulation results. This is even true for systems close to the nucleation threshold: the system with $n=30$ corresponds to the last point of full nucleation control in figure~\ref{FIG:NewDataXn}. Also in analogy to the striped pattern, the onset of nucleation can be seen from the increase in concentration toward the corner of the pattern close to $\dimless{r}=1/\sqrt{2}$. Both cases exhibit a maximal concentration significantly lower than the $F/g$ limit, which in dimensionless units is given by $2/\beta^2=1.38$ and $0.61$ for $n=20$ and $30$, respectively. These findings correspond to the picture that the influence of desorption becomes less important with decreasing $p$ and increasing $s$.

\begin{figure}[h!]
%\subfigure[\label{FIG:NewDataProbability}Nucleation probability. ]{
\includegraphics[width=\linewidth]{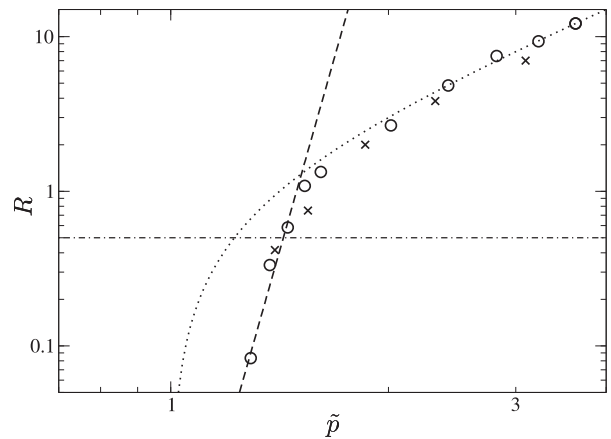}
%}
\caption{{\label{FIG:NewDataR}}Nucleation probability in dependence of the reduced periodicity for the fluxes $1.43\cdot10^{-7}\,\mrm{pt}/(\mrm{step}\,\sigma^2)$ ($\circ$) and $1.43\cdot10^{-6}\,\mrm{pt}/(\mrm{step}\,\sigma^2)$ ($\times$). As before the limit behavior for $\tilde{p}\ll 1$ and $\tilde{p}\gg 1$ is indicated by dotted lines. The dashed-dotted line marks $\xn=1/2$. The dashed lines symbolize represents the estimated expectation in the limit $R\propto F^d$ according to equation~\eqref{EQ:Deff} using $f\approx16.5$.}
\end{figure}

We further attempt to examine the flux dependence of $\tilde{p}$ in the described system. To make use of the previously presented numerical analysis (see especially figure~\ref{FIG:DNumerical}(b)), an estimate of $\istar$ is needed. According to the Walton relation all unstable clusters are expected to be in equilibrium while stable clusters continue to grow, thus $\istar$ can be extracted from the distribution of the cluster sizes. This in combination with visual evaluation leads to an estimate of $\istar\approx5$ for the described system. As can be seen from figure~\ref{FIG:NewDataXn}, the regime of nucleation control with $R\ll 1$ is found up to $\tilde{p}\approx 1.3$ which in combination with the above $\lambda$ and $s_\mrm{eff}$ lead to a reduced dot radius of $\dimless{s}_\mrm{eff}\approx 0.18$. At this $\dimless{s}_\mrm{eff}$ and $\istar$ figure~\ref{FIG:DNumerical}(b) (which with $\beta=1.9$ corresponds to the present case) exhibits $b \approx 0.05$ and consequently one expects only a slight $F$ dependence of $\tilde{p}$.

With respect to a change in $F$ the Monte Carlo system is limited in two ways: on the one hand, a sufficient reduction of $F$ requires an according increase in substrate size and particle numbers which in turn leads to a significant increase in computational effort; while a corresponding increase of $F$ on the other hand brings the system to its resolution limit, when the observed distances between islands reduce to the scale of only a few atoms.

Based on these considerations a series of simulations with a tenfold increased flux of $F=1.43\cdot10^{-6}\,\mrm{pt}/(\mrm{st}\,\sigma^2)$ was conducted, the results of which are represented by the crosses in figure~\ref{FIG:NewDataXn}. Similar to before, the three systems following the loss of nucleation control were determined by twelve independent runs, while the other were obtained from six; the nucleus densities were determined at the beginning of the saturated regime before the onset of coalescence. In contrast to the previous case however, this time frame is not as wide as before due to the increased flux. This leads to a problem, because the scaling relations~\eqref{EQ:UnpatternedScaling} and~\eqref{EQ:Defa} are based on systems at the same coverage, \emph{not} time; hence, for an adequate comparison the current simulations would have to be evaluated at $3\cdot10^4$ steps, a time at which saturation has not set in, yet. An evaluation of both systems at a larger coverage is not possible, because of the onset of coalescence. This incompatibility is substantiated by an evaluation of the nucleus densities. At the increased flux, the surface exhibits a higher island density leading to $\lambda=12.9\,\sigma$, whereas one would ideally expect a value of $10.9\,\sigma$ according to equations~\eqref{EQ:UnpatternedScaling} and~\eqref{EQ:DefChi} with $\chi=5/7$. Additionally, this might also be attributed to the comparatively larger size of the stable nucleus in relation to $\lambda$ which may promote coalescence effects reducing the number of nuclei relative to the theoretical expectation and thus resulting in an increased value of $\lambda$. With respect to the loss of nucleation control the obtained data sets do not exhibit any particular difference. A further evaluation based on the estimated value of $b$ is not possible, however, as the data could not be obtained at the same coverage.

As an alternative means of comparison between the nucleation control of theory and simulation we turn to mean number of additional nuclei $R$, which can be obtained from $\xn$ via equation~\eqref{EQ:DefXn} and which is depicted in figure~\ref{FIG:NewDataR}. Its $\tilde{p}$ dependency according to equation~\eqref{EQ:Deff} can be estimated in the following way: taking, as before, $\dimless{s}\approx 0.18$ for $\tilde{p}\approx 1.3$, the numerical evaluation leading to figure~\ref{FIG:DNumerical}(b) yields $a\approx -3.3$. From this one can in turn find $f\approx16.5$ by means of equation~\eqref{EQ:DeffSecond} with $\istar=5$. Recalling its definition~\eqref{EQ:Deff}, $f$ is an expression for the slope in figure~\ref{FIG:NewDataR}. As can be seen from the dashed line, the observed behavior corresponds nicely to the theoretical estimation, substantiating the validity of the conducted ansatz.

\section{Discussion and Summary}
In this paper we examined the loss of nucleation control in epitaxial growth on patterned substrates with special emphasis on the pattern spacing. Traditionally the transition from (nearly) full to partial nucleation control is believed to occur when the nucleus density of the pattern becomes lower than the density on an unpatterned surface (which in a square pattern corresponds to $\tilde{p}=1$). This is just a first approximation, however. For a more detailed understanding we analyzed the formation of the first nucleus by means of theory and simulation.

We conducted continuous Monte Carlo simulations of one and two-dimensional patterns and compared these results to the theoretical predictions, that are based on non-interacting adatoms in the limit of a stationary adatom concentration field. In both cases the obtained concentration profiles agree well with the predictions, justifying the chosen limits which we show are always accessible at a sufficiently low flux and sufficiently long times. Only close to the nucleation of a new dot the mutual stabilization of the adatoms becomes relevant. 

Based on the stationary concentration and in principle similar to nucleation theory \cite{VenablesPhilosMag,IslandsMoundsAtoms,EvansThielBarteltReview,MaassReview} we derived an expression for the number of additional nuclei ($R$) formed aside of the pattern by statistical fluctuation. We then discuss the scaling behavior ($b$) of flux ($F$) and pattern spacing ($p$) at constant $R$, i.e.\ $\xn$, by numerical methods complemented by analytical evaluations of some accessible limit cases. \ed{The simulated nucleation behavior stands in good agreement with these results: first, as predicted}, the point where nucleation control is lost does not exhibit any significant dependency on the flux; and second the expected increase of $R$ with $\tilde{p}$ agrees very well with the simulation data. All these observations support the validity of the proposed model.

The present approach, however, does possess certain limitations: (a) it is based on the assumption of static adatom concentration fields on a patterned surface, meaning the dot density of the pattern has to be similar to the density of a respectively \emph{saturated} unpatterned surface. \ed{We hence require that the experiments are quenched at a time where unpatterned surface is past the regime of rapid nucleation and in the regime of} $\dd N/\dd t\approx 0$. (b) Due to the assumption of the static concentration field the approach is also limited to the regime of $\xn\approx 1$. As soon as more than one additional nucleus is formed, i.e.\ $\xn$ becomes smaller than $1/2$, this assumption is not justified any more. This behavior is also exhibited by the simulations. (c) Finally, the presented approach is not able to explain some behavior observed in Ref.~\onlinecite{ChiPRL} which exhibited a significant loss of nucleation control for $\tilde{p}<1$. On the one hand, this suffered from a lack of experimental data concerning $D,g,$ and $F$. On the other hand the inclusion of dynamic effects in a system which has not yet reached the steady state might be necessary.

Therefore, the next step is to generalize the concentration field approach and to take the formation of more than one additional nucleus into account. This can be done e.g.\ by means of a mean field simulation in which the discretized concentration field is propagated in time and new nuclei are introduced in the sense of a level set method. Another important question deals with the geometric shape of the clusters. Here comparison with free energy minimizations on the basis of density field theory \cite{Bauer1,Bauer2} may yield interesting insight.

\section*{Acknowledgments}
The authors would like to acknowledge fruitful discussions with \name{J. Krug}, \name{P. Maass}, as well as \name{L. F. Chi} and her group. F.~K. thanks the ``Fonds der Chemischen Industrie" and the ``NRW Graduate School of Chemistry'' for financial support. J.~Z. is grateful for support by DFG grant SFB~458.

\begin{appendix}
\section{\label{HL:APPStationaryConcRegime}Stationary Concentration Regime}
The following considerations illustrate the limit of the stationary concentration field, which provides the basis for the calculation of the remaining nucleation probability in section~\ref{HL:Theory}. Starting with the one-dimensional case and with $g=0$ for simplicity one can express the time dependent concentration by means of the limit concentration $c(x)_\infty$ as given by equation~\eqref{EQ:Conc1Deq} in combination with $\Delta c(x,t)$ describing the difference to $c(x)_\infty$:
\begin{equation}
c(x,t) = c(x)_\infty + \Delta c(x,t)
\,.
\end{equation}
The difference can then be expanded by a \name{Fourier} series which, having to fulfill the diffusion equation~\eqref{EQ:DiffusionEquation} as well as $\Delta c(0,t)=\Delta c(p,t)=0$ turns out to be
\begin{equation}
\Delta c(x,t) = \sum_n a_n\, \Sin{\frac{n\pi}{p}x} \, \Exp{-D\frac{n^2\pi^2}{p^2}\,t}
\,.
\end{equation}
The \name{Fourier} coefficients are determined from the initial condition $\Delta c(x,t) = -c(x)_\infty$. As the higher modes decay with $n^2$ only the $n=1$ mode with
\begin{equation}
a_1 = - \frac{F}{D}\,\frac{4\,p^2}{\pi^3}
\end{equation}
dominates the long-time approach of $c(x)_\infty$. Concerning the exponential decay of $\Delta c(x,t)$ one can assume the stationary state to be reached when
\begin{equation}\label{EQ:StationaryTime}
t > \frac{p^2}{D\pi^2}.
\end{equation}

For the model proposed in section~\ref{HL:Theory} to hold the considered system has to fulfill two requirements: first, to reach a steady state the time has to be larger than a critical time $t_\mrm{c}$ (given e.g.\ by equation~\eqref{EQ:StationaryTime}). This leads to a general, monotonously growing dependence of $t_\mrm{c}(p)$ and its respective inverse $p_\mrm{c}(t)$. Second, to provide nearly full nucleation control we require a fixed nucleation probability in the range $0 < R \ll 1$, on the basis of which the critical flux can be expressed by the general relation $F_\mrm{c}(p,t,R)$, which decreases monotonously with $p$ and $t$. It should be noted that these critical parameters are not to be mistaken for the ones introduced in section~\ref{HL:Theory}. In the deterministic limit of $\istar\rightarrow\infty$ at $g=0$ the nucleation probability is directly proportional to the maximal concentration from equation~\eqref{EQ:Conc1DMax} and one obtains $F_\mrm{c}\propto p^{-2}$.

\begin{figure}[h]
\subfigure[\label{FIG:tConst}\ $t=\mrm{const}$]{
\includegraphics[width=0.45\linewidth]{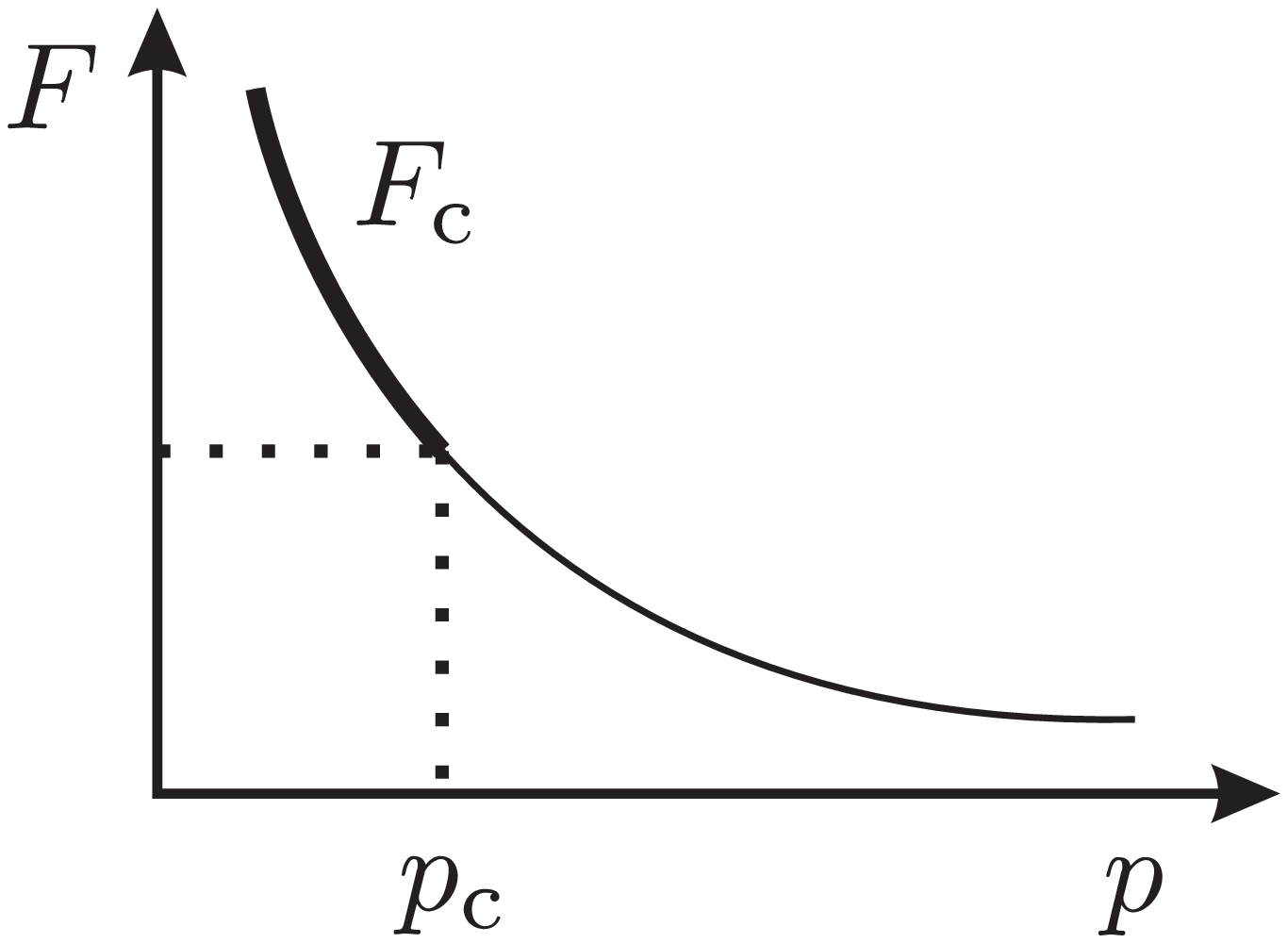}%
}\hfill
\subfigure[\label{FIG:pConst}\ $p=\mrm{const}$]{
\includegraphics[width=0.45\linewidth]{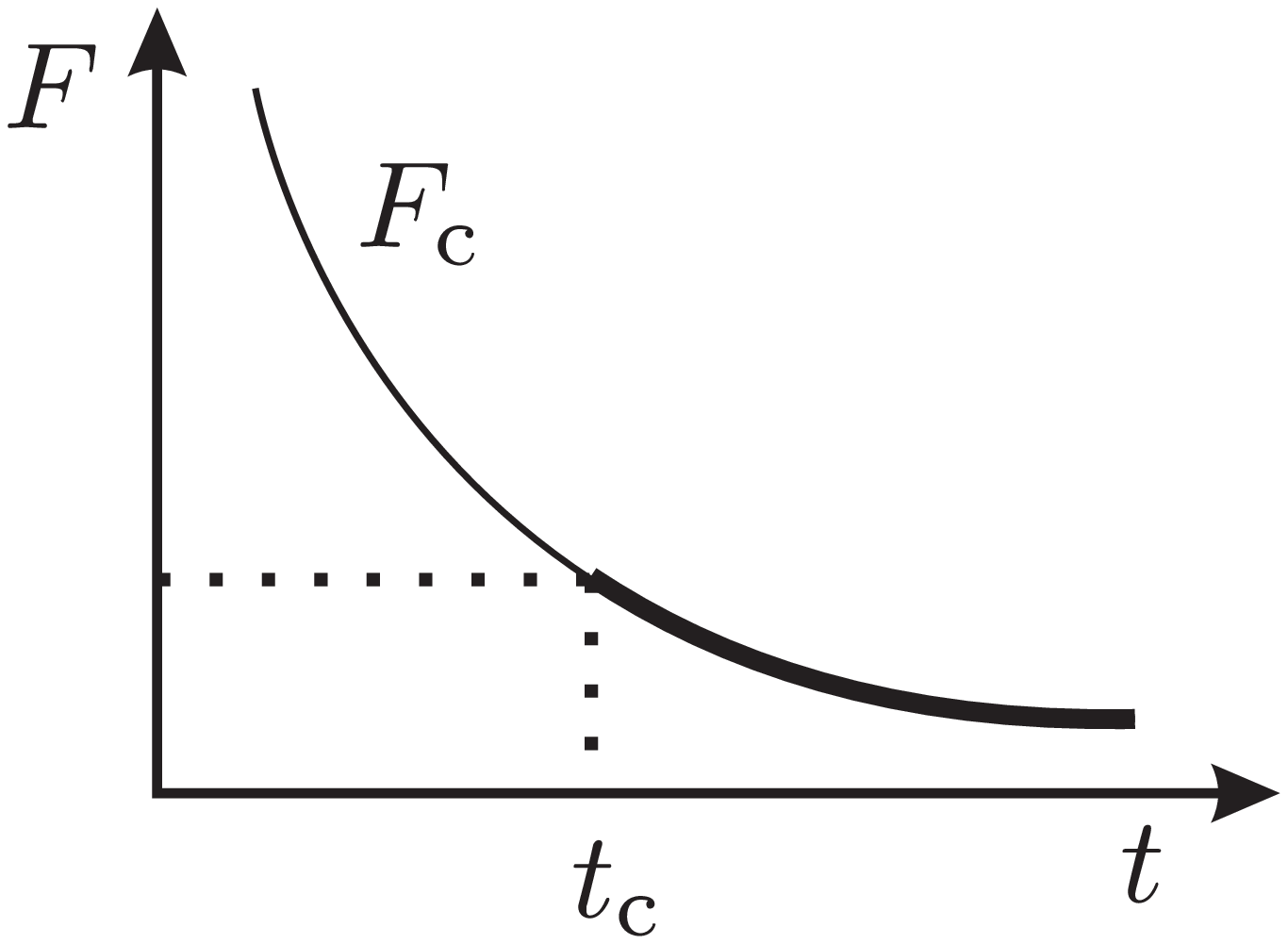}%
}
\caption{\label{FIG:Sketches}Illustrations for the qualitative examination of $F_\mrm{c}$}
\end{figure}

Finally one can discuss two general scenarios: (a) at fixed time scale $t$ the stationary state will be reached for all $p<p_\mrm{c}(t)$. The second requirement of a fixed $R$ can then only be satisfied by a flux of $F>F_\mrm{c}(p_\mrm{c},t,R)$. This is illustrated by the thick solid line in figure~\ref{FIG:tConst}. (b) on the other hand one can consider a scenario at constant $p$. In this case the stationarity requirement is fulfilled if $t>t_\mrm{c}(p)$ while for $R$ to be retained the flux has to be chosen appropriately low satisfying $F<F_\mrm{c}(p,t_\mrm{c},R)$; an illustration of which can be found in figure~\ref{FIG:pConst}. Consequently, the requirements of the proposed model can in principle be satisfied for any experimental setup if the flux is chosen appropriately.

\section{\label{HL:APPStationaryConcNumerics}Numerical Evaluation of the Stationary Concentration}
In this appendix we evaluate the analytical approximation of a radially isotropic concentration field from equations~\eqref{EQ:cisoeqn} and~\eqref{EQ:cmaxeqn} by comparison to the results obtained from numerical solution of the diffusion equation~\eqref{EQ:DiffusionEquation}. The diffusion field is discretized and diffusion is expressed by the finite differences of equation~\eqref{EQ:FiniteDifferences}. Sinks are modeled by a combination of grid points to represent the desired shape.

First, the hexagonal case, being closer to the circular symmetry, is modeled on a $111 \times 192 $ grid with $\Delta x=\Delta y=1~\mrm{u}$ approximating the side relation of $X/Y=1/\sqrt{3}$ at $F/D=40\,\mrm{pt/\mrm{u}4}$ and $g=0/\mrm{t}$. A sink is placed at the corner and in the center of the unit grid. In combination with periodic boundaries this produces the desired hexagonal lattice. Three different shapes of sinks are realized: circles with radius $s$, squares with edge length $2\,s$ and equilateral triangles with edge length $2\,s$.

\begin{figure}
%\subfigure[\label{FIG:heqnhex}Hexagonal Lattice]{
\includegraphics[width=\linewidth]{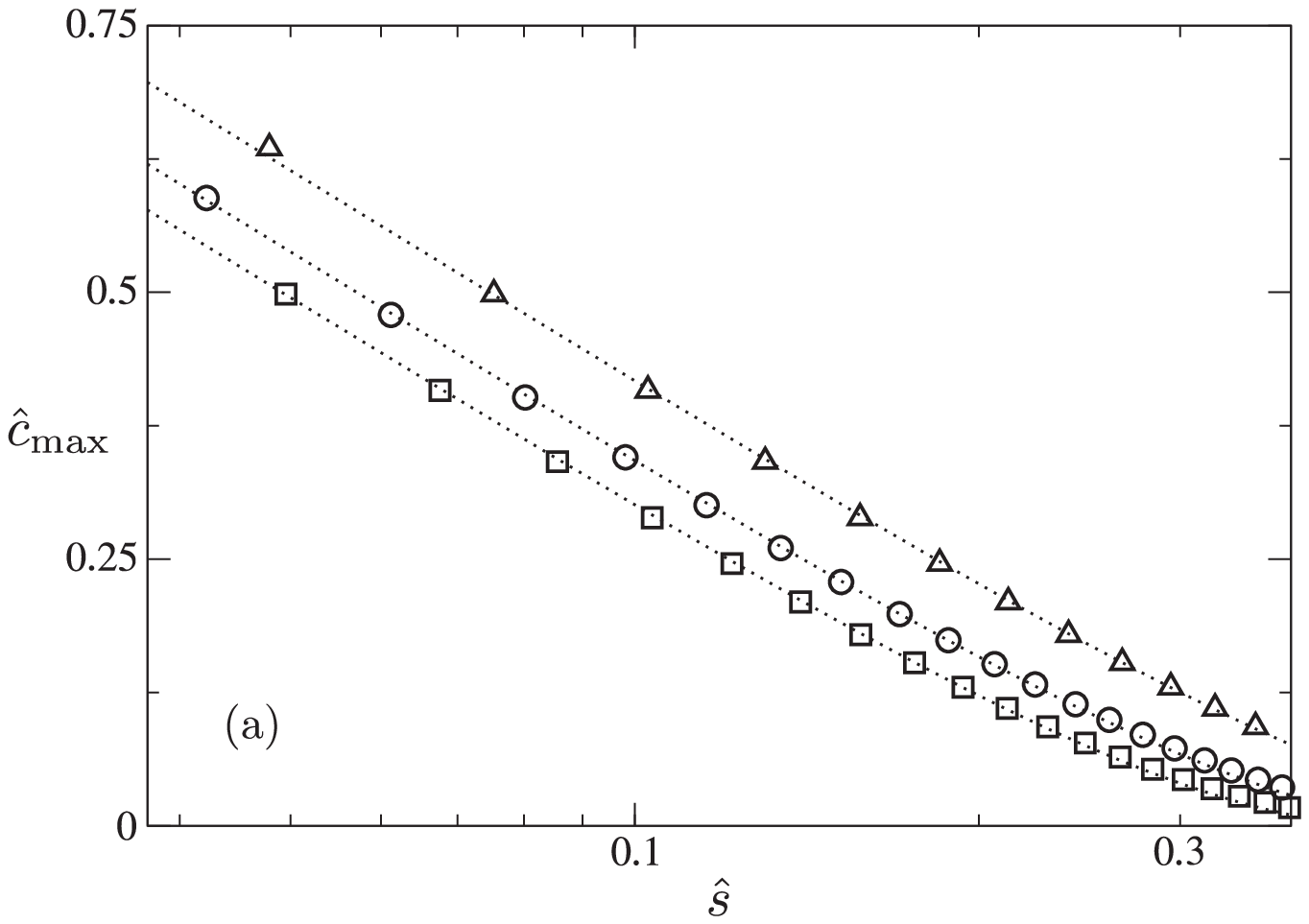}%
%}
\par
%\subfigure[\label{FIG:heqnsq}Squared Lattice]{
\includegraphics[width=\linewidth]{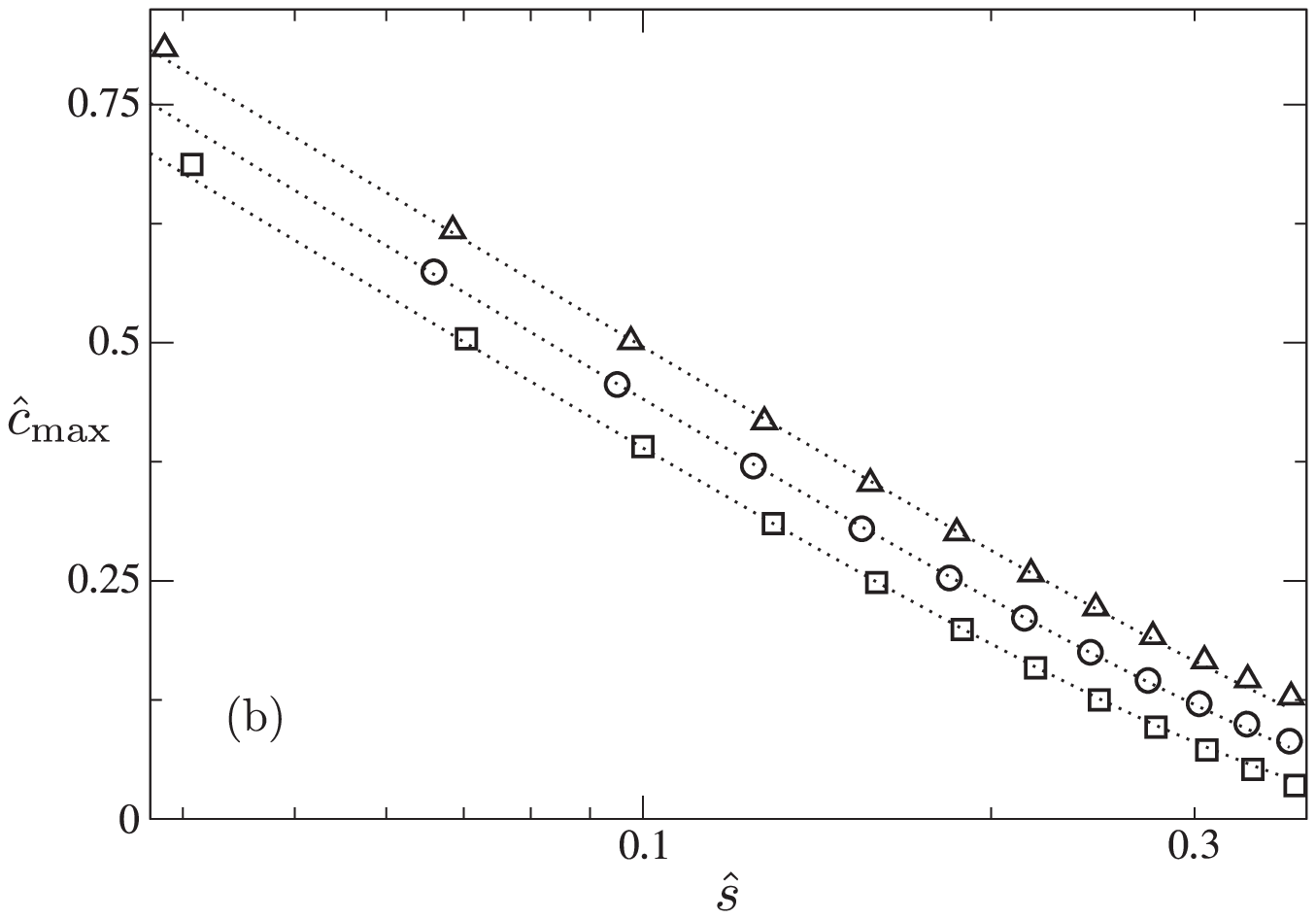}%
%}
\caption{\label{FIG:heqn}Numerically determined $\dimless{c}_\mrm{max}(\dimless{s})$ for circular, squared and triangular dots on hexagonal (a) and squared (b) lattices. Dot shapes are indicated by the symbols. The dotted lines are fitting functions according to equation~\eqref{EQ:hfiteqn} with the respective parameters from table~\ref{TAB:hfitpars}.}
\end{figure}

The obtained $\dimless{c}_\mrm{max}(\dimless{s})$ are depicted in figure~\ref{FIG:heqn}(a). According to the theoretical prediction of equation~\eqref{EQ:cmaxeqn}, the data is fitted using
\begin{equation}\label{EQ:hfiteqn}
\dimless{c}^\mrm{iso}_\mrm{max, fit}(\dimless{s}) = A_1 \cdot \frac{1}{4}\left( -\ln\left(2\,A_0\dimless{s}\right) + 2\left(A_0\dimless{s}\right)^2 - \frac{1}{2} \right)
\,,
\end{equation}
with $A_0$ and $A_1$ as fit parameters. Since $A_1$ seems to be independent of the dot shape, it is determined from an average over fits of the same pattern; $A_0$ is then re-calculated based on this result. The obtained fitting parameters are given in table~\ref{TAB:hfitpars}.
\begin{table}
\begin{tabular}{l|ccc|ccc|}
pattern: & \multicolumn{3}{c|}{hexagonal} & \multicolumn{3}{c|}{quadratic}\\
dot shape:     &  circle & square & triangle    & circle & square & triangle \\\hline
$A_0$    &   0.938 &   1.09 & 0.718       & 0.774 & 0.913  & 0.650 \\
%$A_1$    & \multicolumn{3}{c|}{1.15}      &  \multicolumn{3}{c|}{1.28}     \\ 
$A_1$    &   1.15  &   1.15 & 1.15        & 1.28  & 1.28  & 1.28    \\
$B$      &   $\sqrt{3}/2$  &   $\sqrt{3}/2$ & $\sqrt{3}/2$        & $1/\sqrt{2}$  & $1/\sqrt{2}$  & $1/\sqrt{2}$    \\
%a & b & c & d & e \\
%f & g & h & i & j \\
\end{tabular}
\caption{\label{TAB:hfitpars}Fitting parameters for equations~\eqref{EQ:hfiteqn} and~\eqref{EQ:cfiteqn} for different patterns and island morphologies. $A_0$ and $A_1$ are determined based on numerical solution. $B$ is derived geometrically.}
\end{table}

The scaling parameter $A_0$ can be interpreted in the sense of an effective radius to approximate a non-circular dot by a circle: assuming that for reasonably small $\dimless{s}$ the influence of a non-circular dot diminishes rather quickly as one moves away from it, an approximately radially isotropic profile is obtained; this could in turn just as well have been caused by a circular dot of an appropriate effective radius. Adopting the picture that primarily the area of the dot has to be constant, one can estimate this effective radius from simple geometrical considerations. For the quadratic and triangular shapes one would expect the radii to be related to the circular radius by factors of $(\pi/4)^{1/2}$ and $(\pi/\sqrt{3})^{1/2}$, respectively. This corresponds well to the ratios of the according $A_0$ from table~\ref{TAB:hfitpars}.

A square lattice is realized under the same conditions as above with the exception of a $135 \times 135 $ grid. The adoption of a radially isotropic concentration is not well suited for this lattice; nevertheless, the obtained data can be properly fitted according to equation~\eqref{EQ:hfiteqn} as shown in figure~\ref{FIG:heqn}(b). The parameters were determined as before and are given in table~\ref{TAB:hfitpars}. Again, the data of all calculated shapes can be expressed by the same $A_1$, but in contrast to the hexagonal lattice the ratios of the $A_0$ do not reflect the geometrically expected effective radii as closely as before.

\begin{figure}[b]
\includegraphics[width=\linewidth]{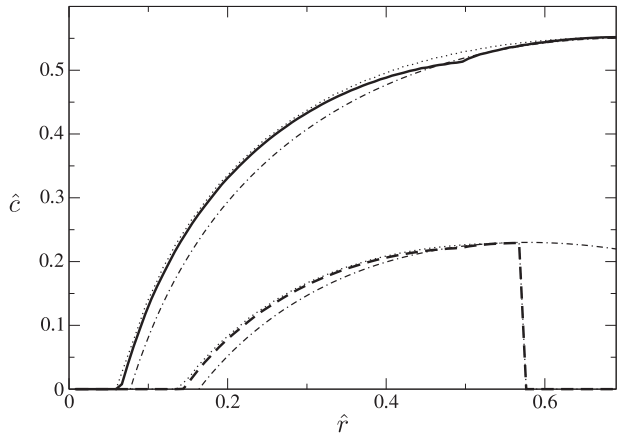}
\caption{\label{FIG:2Dapprox} The radial concentration profiles for a circular dot with $\dimless{s}=0.07$ on a cubic lattice (solid line) and $\dimless{s}=0.15$ in a hexagonal lattice (dashed line). The bold and dotted lines are obtained by numerical simulation and symbolize the mean and maximal concentration, respectively. The dashed-dotted lines are the approximations according to equation~\eqref{EQ:cfiteqn}.}
\end{figure}

Incorporating $A_0$ and $A_1$ into the radial concentration~\eqref{EQ:cisoeqn} one obtains
\begin{equation}\label{EQ:cfiteqn}
\dimless{c}^\mrm{iso}(\dimless{r},\dimless{s}) = A_1 \left[ \frac{1}{2}\left((A_0\dimless{s})^2 - (B\dimless{r})^2 \right) + \frac{1}{4}\Ln{\frac{B}{A_0}\frac{\dimless{r}}{\dimless{s}}}\right]
,
\end{equation}
where the introduction of the additional parameter $B$ rescales the space coordinate such that the maximum is moved from $\dimless{r}=1/2$ to the maximal distance of the respective lattice. $B$ can be derived from simple geometrical considerations and is also listed in table~\ref{TAB:hfitpars}. Exemplary curves for a quadratic dot of $\dimless{s}=0.07$ and a circular one with $\dimless{s}=0.15$ are depicted in figure~\ref{FIG:2Dapprox} as dashed-dotted lines in comparison to the numerically determined profiles. The numerical results (bold lines) were radially averaged over one unit-cell leading to an edge at $\dimless{r}=1/2$, where the boundary of the cell is met. This edge is more pronounced for the quadratic lattice due to its less circular shape. As expected, the maximal concentrations (dotted lines) do not exhibit this edge. It can also be seen that equation~\eqref{EQ:cfiteqn} only closely approximates the latter part of the profiles as $A_0$ and $B$ do not sufficiently compensate to reproduce the used $\dimless{s}$.

\end{appendix}

\vfill
\bibliography{/nfs/g/felixk/Dokumente/Literatur/Literatur}
%\bibliography{Literatur}

\end{document}